\pdfoutput=1

\documentclass[a4paper,fleqn,usenatbib]{mnras}

\usepackage{newtxtext,newtxmath}
\usepackage{xcolor}
\usepackage{ulem}

\usepackage[T1]{fontenc}
\usepackage{ae,aecompl}

\usepackage{graphicx}	
\usepackage{amsmath}	

\newcommand{\nudot}{\dot{\nu}}
\newcommand{\Tobs}{T_{\rm obs}}

\newcommand{\vref}{_{\rm ref}}
\newcommand{\Ng}{N_\mathrm{g}}
\newcommand{\Np}[1]{174}
\newcommand{\Nz}[1]{233}

\newcommand{\logBABglitch}{3.66\pm0.19}
\newcommand{\logBACglitch}{13.93\pm0.17}
\newcommand{\logBADglitch}{89.91\pm0.17}
\newcommand{\logBBCglitch}{10.27\pm0.19}
\newcommand{\logBBDglitch}{86.25\pm0.19}
\newcommand{\logBCDglitch}{75.98\pm0.18}

\newcommand{\logBABall}{3.36\pm0.19}
\newcommand{\logBACall}{18.11\pm0.18}
\newcommand{\logBADall}{128.90\pm0.18}
\newcommand{\logBBCall}{14.75\pm0.19}
\newcommand{\logBBDall}{125.53\pm0.19}
\newcommand{\logBCDall}{110.78\pm0.18}

\newcommand{\ModelAAall}{0.0099_{-0.003}^{+0.004}}
\newcommand{\ModelBAall}{0.013_{-0.01}^{+0.04}}
\newcommand{\ModelCAall}{6.5{\rm e-}05_{-1{\rm e-}05}^{+1{\rm e-}05}}
\newcommand{\ModelDAall}{2.9{\rm e-}06_{-2{\rm e-}06}^{+6{\rm e-}06}}

\newcommand{\ModelAAglitch}{0.0066_{-0.002}^{+0.003}}
\newcommand{\ModelBAglitch}{0.0082_{-0.006}^{+0.02}}
\newcommand{\ModelCAglitch}{8.8{\rm e-}05_{-1{\rm e-}05}^{+2{\rm e-}05}}
\newcommand{\ModelDAglitch}{9.3{\rm e-}06_{-6{\rm e-}06}^{+2{\rm e-}05}}

\newcommand{\ModelAgammaall}{0.31_{-0.03}^{+0.03}}
\newcommand{\ModelBalphaall}{-0.31_{-0.08}^{+0.08}}
\newcommand{\ModelBbetaall}{0.31_{-0.03}^{+0.03}}
\newcommand{\ModelCbetaall}{0.23_{-0.02}^{+0.02}}
\newcommand{\ModelDalphaall}{0.24_{-0.06}^{+0.06}}

\newcommand{\ModelAgammaglitch}{0.27_{-0.03}^{+0.03}}
\newcommand{\ModelBalphaglitch}{-0.27_{-0.08}^{+0.08}}
\newcommand{\ModelBbetaglitch}{0.27_{-0.03}^{+0.03}}
\newcommand{\ModelCbetaglitch}{0.19_{-0.02}^{+0.02}}
\newcommand{\ModelDalphaglitch}{0.18_{-0.06}^{+0.06}}

\title[Glitch rate law inferred from radio pulsars]{An updated glitch rate law inferred from radio pulsars}

\author[Millhouse et al.]{
M. Millhouse,$^{1,2}$ \thanks{E-mail: meg.millhouse@unimelb.edu.au}
A. Melatos, $^{1,2}$
G. Howitt,$^{1,2}$
J.B. Carlin,$^{1,2}$
L. Dunn,$^{1,2}$
G. Ashton$^{1,3}$
\\
$^{1}$ Australia Research Council Centre for Excellence for Gravitational Wave Discovery (OzGrav)\\
$^{2}$ School of Physics, University of Melbourne, Parkville, VIC 3010, Australia\\
$^{3}$ School of Physics and Astronomy, Monash University, VIC 3800, Australia
}

\date{Accepted XXX. Received YYY; in original form ZZZ}

\pubyear{2022}

\begin{document}
\label{firstpage}
\pagerange{\pageref{firstpage}--\pageref{lastpage}}
\maketitle

\begin{abstract}
Radio pulsar glitches probe far-from-equilibrium processes involving stress accumulation and relaxation in neutron star interiors.  Previous studies of glitch rates have focused on individual pulsars with as many recorded glitches as possible.  In this work we analyze glitch rates using all available data including objects that have glitched never or once.  We assume the glitch rate follows a homogeneous Poisson process, and therefore exclude pulsars which exhibit quasiperiodic glitching behavior. Calculating relevant Bayes factors shows that a model in which the glitch rate $\lambda$ scales as a power of the characteristic age $\tau$ is preferred over models which depend arbitrarily on powers of the spin frequency $\nu$ and/or its time derivative $\dot{\nu}$.  For $\lambda = A (\tau/\tau\vref)^{-\gamma}$, where $\tau_{\rm ref}=1\ {\rm yr}$ is a reference time, the posterior distributions are unimodal with $A=\ModelAAglitch\ \rm{yr}^{-1}$, and $\gamma=\ModelAgammaglitch$.
Importantly, the data exclude with 99\% confidence the possibility $\gamma=1$ canvassed in the literature.
When objects with zero recorded glitches are included, the age-based rate law is still preferred and the posteriors change to give $A=\ModelAAall\ \rm{yr}^{-1}$, and $\gamma=\ModelAgammaall$.  The updated estimates still support increased glitch activity for younger pulsars, while demonstrating that the large number of objects with zero glitches contain important statistical information about the rate, provided that they are part of the same population as opposed to a disjoint population which never glitches for some unknown physical reason. \end{abstract}




\section{Introduction}
Electromagnetic observations of pulsars have shown that neutron stars spin down secularly, with occasional rapid spin-up events known as ``glitches".  The glitch trigger mechanism is unknown, but one possibility involves a stick-slip, avalanche-mediated interaction between the rigid crust and superfluid core of the star~\citep{1975Natur.256...25A,1984ApJ...276..325A,Melatos_2008}.  Understanding the pulsar glitch mechanism will furnish insights into nuclear physics~\citep{2018IAUS..337..203H,2015IJMPD..2430008H}.  Traditional glitch finding methods analyze pulse times of arrival (ToA) to detect step-like departures from a secular phase model~\citep{Espinoza2011}.  More recent glitch finding methods use Bayesian model selection~\citep{10.1093/mnras/stt2122,10.1093/mnras/stx702,Lower_2020,2021MNRAS.tmp.2433L} and hidden Markov models~\citep{2020ApJ...896...78M,2021MNRAS.tmp.2433L}.
To date, nearly 200 pulsars have been observed to glitch~\citep{2005yCat.7245....0M,2005AJ....129.1993M,Espinoza2011}, with a total of over 500 events.  The event rate varies significantly between pulsars. While some pulsars glitch every year or so, others have only recorded one glitch in the past 50 years. Many pulsars have not been observed to glitch at all.

Long-term glitch statistics have been studied in individual objects with a relatively high number of observed events~\citep{Melatos_2008,Espinoza2011,2017PhRvD..96f3004A,2018ApJ...863..196M,Howitt_2018,2019A&A...630A.115F,2021Univ....7....8M}, as well as the radio pulsar population as a whole~\citep{2000MNRAS.315..534L,2017A&A...608A.131F,2019RAA....19...89E}.  While the investigations into individual pulsars tend to look for relations between glitch sizes (the fractional spin frequency jump $\Delta \nu/\nu$) and the waiting times $\Delta t$ between the events,  the population studies of pulsars look at how glitch activity relates to stellar properties such as age, magnetic field, or spin-down rate.  
Understandably much of the focus has been on objects that have glitched multiple times.  This is appropriate for certain analyses, specifically those aiming to construct probability density functions (PDFs) or cross- and autocorrelation statistics of $\Delta \nu$ and $\Delta t$~\citep{2019MNRAS.488.4890C} and those trying to estimate properties of specific neutron stars~\citep{2020MNRAS.492.4837M}.  However, pulsars with one or even zero glitches, which represent a large percentage of the population, carry important information about glitch activity, especially the glitch rate per pulsar. Even pulsars which have zero recorded glitches to date can lend insight into global pulsar properties~\citep{2017A&A...608A.131F}.

In this work, we relate glitch rate $\lambda$ to three measurable properties of pulsars: the spin frequency $\nu$, the spin-down rate $\dot{\nu}$, and the characteristic spin-down age $\tau = \nu/(2\left|\dot{\nu}\right|)$ within a Bayesian framework. Our analysis builds on extensive previous work~\citep{McKenna1990,2000MNRAS.315..534L,Espinoza2011,2018MNRAS.473.1644A,2020MNRAS.498.4605H}. Most recently, ~\cite{2017A&A...608A.131F} binned objects by the glitch activity parameter~\citep{McKenna1990} and found correlations between the glitch activity and $\nudot$, $\tau$, as well as the spin-down luminosity $\propto \nu \dot{\nu}$.  Intriguingly they found evidence for an empirical relation of the form $\lambda \propto \dot{\nu}$ dominated by glitches above a certain size; see Section 4.2 and Figure 6 in~\citet{2017A&A...608A.131F}.

This paper is complementary to previous studies but differs from them in two ways: (i) it includes all pulsars, including those which have glitched once or not at all, and including those with small glitches, even though samples of small glitches may be incomplete; and (ii)  it selects formally between phenomenological glitch rate models within a Bayesian framework, extending previous work involving binning, least-squares fits, and summary statistics like the Akaike information criterion.

We focus on the glitch rate and leave the inclusion of glitch sizes to future work. Specifically, we use all the data contained in the Jodrell Bank Observatory\footnote{\url{http://www.jb.man.ac.uk/pulsar/glitches.html}} and Australia Telescope National Facility\footnote{\url{https://www.atnf.csiro.au/research/pulsar/psrcat/}} glitch catalogues~\citep{2005AJ....129.1993M,2005yCat.7245....0M,Espinoza2011,2021MNRAS.tmp.3056B}, including pulsars which have glitched once.  We then redo the analysis with radio pulsars that have never glitched to test the effect on the inferred glitch rate.  The latter test assumes that all known pulsars glitch eventually, once enough time elapses. It does not apply if the non-glitching pulsars are a disjoint population which never glitch for some unknown physical reason, and is eminently possible of course. 

The paper is organized as follows. In Section~\ref{Sec:dataset} we discuss the pulsars in the data set (both glitching and non-glitching).  In Section~\ref{Sec:rates} we produce posterior distributions of the glitch rate per pulsar assuming that every pulsar glitches according to a homogeneous (i.e. constant-rate) Poisson process. Section~\ref{Sec:Models} introduces four phenomenological population models that relate the glitch rate to $\nu$, $\dot{\nu}$, and $\tau$. Sections~\ref{Sec:GlitchResults} and~\ref{Sec:NonglitchResults} presents the results of Bayesian model selection and parameter estimation applied to the four rate laws, with and without non-glitching pulsars respectively.
Section~\ref{Sec:Biases} discusses some of the factors that bias our parameter estimation, including the observation times of the pulsars and the completeness of existing glitch catalogs.
The physical implications for the glitch mechanism are discussed briefly in Section~\ref{Sec:conclusion}.

\section{Data}\label{Sec:dataset}
In this work we analyse pulsars that have been observed to glitch at least once, as well as pulsars that have been monitored for glitches without any being recorded. Here we describe these two data subsets.

\subsection{Objects with at least one recorded glitch}\label{Sec:DataGlitchers}
In this work, we analyse \Np{} pulsars that have been observed to have glitched at least once over the last 50 years~\citep{2005yCat.7245....0M,2005AJ....129.1993M,Espinoza2011,2021MNRAS.tmp.3056B}.  The majority of these pulsars have a small number of observed glitches, $\Ng$, with 101 objects having only glitched once during the observation period.  The object included in this analysis with the most glitches recorded is PSR J1740-3015 with 36 glitches\footnote{PSR J0537$-$6910, with $\Ng=45$ glitches periodically and is excluded form the analysis; see Table~\ref{Tab:PoissonGlitchers}. In addition to the 45 glitches in the Jodrell Bank glitch catalogue, an analysis of x-ray data from NICER has reported eight new glitches~\citep{2020MNRAS.498.4605H}.}.
  
 The models considered in this work require measurements of $\nu$ and $\nudot$, and so we only include objects with estimates for these values provided by the Australian Telescope National Facility Pulsar Catalogue.
In order to estimate the glitch rate of each pulsar, we also need the time period over which it has been observed, $T_\mathrm{obs}$.
We take as the start of $\Tobs$ the date of the discovery publication as reported in the ATNF catalog, and we take MJD 58849 (the beginning of calendar year 2020) as the end of the observational period.  The above definition of $\Tobs$ is an approximation for most pulsars, because monitoring is not always continuous between the start and end dates, and some pulsars are not monitored regularly before the discovery of a first glitch. The consequences of this approximation, which is unavoidable due to gaps in the literature regarding monitoring history, are explored in Section~\ref{Sec:Biases}.
  
  Table~\ref{Tab:rates} lists every glitching pulsar analyzed in this paper, as well as their respective number of glitches observed, observation time, and estimated rate (see Section~\ref{Sec:rates}).
  
In this paper, we interpret rates in terms of a homogeneous Poisson process, following the usual practice in the literature~\citep{Melatos_2008,2017A&A...608A.131F,2021MNRAS.tmp.2433L}. We therefore exclude PSR J0835-4510 (the Vela Pulsar), PSR J0537-6910, and PSR J1341-6220, which exhibit strong evidence for quasiperiodic glitch activity, inconsistent with a homogeneous Poisson process.
Quasiperiodic glitch activity has been noted since the early days of radio pulsar timing. PSR J0835-4510 (the Vela Pulsar) is recognized to glitch strongly once every $\sim 3\,{\rm yr}$~\citep{2013ATel.5406....1B,2008AIPC..983..145B}, although there seems to be a subset of weaker glitches which may or may not recur regularly. PSR J0537-6910 glitches roughly three times per year 
  and displays a strong correlation between glitch size and forward waiting time, which can be used to predict glitch epochs accurately --- the only pulsar discovered until now, where this is possible~\citep{2006ApJ...652.1531M,2018ApJ...863..196M,2018ApJ...852..123F,2018MNRAS.473.1644A}. 
 Recently, ~\cite{Howitt_2018} showed that PSR J1341-6220 exhibits quasiperiodic activity, while it had previously been classified as  Poisson-like~\citep{Melatos_2008}. 
 Statistical analyses of waiting time distributions confirm the existence of quasiperiodic activity, e.g. using the Kolmogorov-Smirnov (KS) test~\citep{Melatos_2008}, kernel density estimator (KDE)~\citep{Howitt_2018}, or Akaike information criterion~\citep{2019A&A...630A.115F}. The PDF is unimodal and peaks at some nonzero waiting time, unlike for Poisson activity, where the PDF is exponential and peaks at zero waiting time. 
 Table~\ref{Tab:PoissonGlitchers} lists the three objects which are identified as exceptions to Poisson activity in the foregoing references, together with the statistical method employed, the number of events, and the raw mean rate $\Ng/T_{\rm obs}$.
 
It is likely that more objects will be added to the quasiperiodic category in the future, as more data are collected. It is also likely that some objects included in the Poisson sample in this paper will turn out to be quasiperiodic, when more data are collected. It is impossible to predict how many there will be, and which ones, so the results of the analysis must be interpreted with this caveat in mind. The quasiperiodic fraction of the population could be as high as $\sim 30\%$ by naive extrapolation, given that three quasiperiodic objects are known out of $\sim 10$ with $\Ng \gtrsim 10$ high enough to be tested~\citep{Howitt_2018,2019A&A...630A.115F}, but it could also be a lot lower, if there is a selection effect at play because quasiperiodic objects have higher $\lambda$ for some physical reason~\citep{2019MNRAS.483.4742C}.

 \begin{table}
\caption{Pulsars that exhibit quasiperiodic glitch activity.  For each pulsar, we list the statistical method that provides evidence for quasiperiodicity, the number of glitches observed, and the raw rate estimate, $\Ng/T_{\rm obs}$.}
\begin{center}
\begin{tabular}{c c c c}
Pulsar & Statistical method & $\Ng$ & $\Ng/T_{\rm obs}$ yr$^{-1}$\\
\hline
J0835-4510 & KS test, Akaike, KDE & 20 & 0.386 \\
J0537-6910 & KS test, Akaike, KDE & 45 & 3.47 \\
J1341-6220 & KDE & 23 & 0.66 \\
\hline
\end{tabular}
\end{center}
\label{Tab:PoissonGlitchers}
\end{table}%

\subsection{Objects with zero recorded glitches}\label{Sec:nonglitchers}
In addition to including data from pulsars with confirmed glitch activity, this work also seeks to investigate the impact of including pulsars with no recorded glitches.  There are a number of ways one could construct this data set. One naive approach is to include all pulsars with $\Ng = 0$.  However there are also compelling reasons not to include every non-glitching pulsar.  Some pulsars which have no associated glitches are not routinely monitored. We do not know if a glitch may have occurred relatively recently; including such infrequently monitored objects in a list of non-glitching pulsars may not be appropriate.  Furthermore, only two millisecond pulsars have ever been observed to have glitched~\citep{McKee2016,2004ApJ...612L.125C}. It is possible that millisecond pulsars belong to a separate population, which glitches infrequently or never due to some unknown physical reason. Equally it is possible that millisecond pulsars glitch just like longer-period pulsars but at slower rates, in line with their greater ages (see Section 4.2 in~\cite{2015ApJ...807..132M}). This issue is revisited in Section~\ref{Sec:completeness} of this paper.
For these reasons, rather than analyzing all pulsars with zero known glitches, we analyze a subset of non-millisecond pulsars that have been routinely observed.

When adding zero-glitch pulsars into the analysis in Section~\ref{Sec:NonglitchResults}, we select pulsars that have been regularly monitored by the UTMOST collaboration using the Molonglo Observatory Synthesis Telescope~\citep{Jankowski2019,Lower_2020,2021MNRAS.tmp.2433L}.  This sample is selected for three reasons: (i) records of $\Tobs$ are more complete than for the pulsar population in general; (ii) UTMOST pulsars are monitored with high cadence and semi-continuously; and (iii) the sample has been searched systematically for glitches down to a certain size. Specifically on point (iii), public data made available by UTMOST are searched for glitches with a hidden Markov model as described in~\cite{2020ApJ...896...78M}. We include the pulsars where no glitches were reported either by~\citet{Lower_2020} or the hidden Markov model search~\citep{Dunn_2020}.
There are a total of \Nz{} pulsars with $\Ng = 0$ included in this data set.  
For these objects, we set the beginning of the observation time as Oct 2015, the start of the UTMOST observing campaign~\citep{Jankowski2019}, and the start of 2020 as the end of the observing time.  For all these pulsars, one has $\Tobs=4.25\ {\rm yr}$.  As in Sec.~\ref{Sec:DataGlitchers}, this definition of $\Tobs$ is an approximation for most pulsars. The observing cadence for these objects ranges from near daily to monthly, with an average of about two weeks.  This is comparable to the observing cadences of approximately 10--14 days for the Jodrell Bank Observatory~\citep{2021MNRAS.tmp.3056B}, and approximately 2--4 weeks for the Parkes radio telescope (also called \textit{Murriyang})~\citep{Yu2013,2021MNRAS.tmp.2433L}.
Table~\ref{Tab:zero_glitchers} lists the pulsars with zero recorded glitches used in this work, their characteristic spin-down age $\tau$, and observation time $T_\mathrm{obs}$.

\section{Poisson glitch rate per object}\label{Sec:rates}

In this section, we estimate the individual glitch rate of every pulsar with at least one recorded glitch. To do so, we assume that glitch activity is a homogeneous (i.e. constant rate) Poisson process over the decades over which pulsars have been monitored. This hypothesis is broadly consistent with observational data for those objects with enough events to justify meaningful statistical tests, e.g. for the existence of an exponential waiting-time PDF~\citep{Wong2001,Melatos_2008,Howitt_2018,2019A&A...630A.115F}. However there are exceptions. 
We exclude PSR J0835-4510 (the Vela Pulsar), PSR J0537-6910, and PSR J1341-6220 which glitch quasiperiodically as discussed in Section~\ref{Sec:dataset}. We also note that PSR J0534+2200, appears to glitch according to an inhomogeneous Poisson process, i.e. the glitch rate appears to change every $\sim 10\,{\rm yr}$ when it is examined closely~\citep{2015MNRAS.446..857L,2019MNRAS.482.3736C}. This pulsar is included in the Poisson analysis in this section as if it glitches at a constant rate, because decadal variations in the glitch rate lie outside the scope of this paper, and the object appears to be in a class of its own for now, given current data.

 If glitch activity is a homogeneous Poisson process, then the glitch rate for an individual pulsar is estimated via the Poisson distribution, viz.
\begin{equation}
  p(\lambda|\Ng,T_\mathrm{obs}) = \frac{(\lambda T_\mathrm{obs})^{\Ng} e^{-\lambda T_\mathrm{obs}}}{\Ng!},
  \label{Eq:Poisson}
\end{equation}
where $\Ng$ is the number of observed glitches, $T_\mathrm{obs}$ is the total amount of time the pulsar has been observed, and $\lambda$ is the glitch rate of that pulsar.  

We use the $\texttt{dynesty}$ nested sampling algorithm~\citep{2020MNRAS.493.3132S} to estimate the posterior of $\lambda$ for each individual pulsar given the data $\Ng$ and $T_\mathrm{obs}$.  For the prior distribution on $\lambda$, we use a log-uniform distribution between $4\times10^{-4}$ yr$^{-1}$ and $40$ yr$^{-1}$. 

\begin{figure}
  \centering
  \scalebox{0.6}{\includegraphics{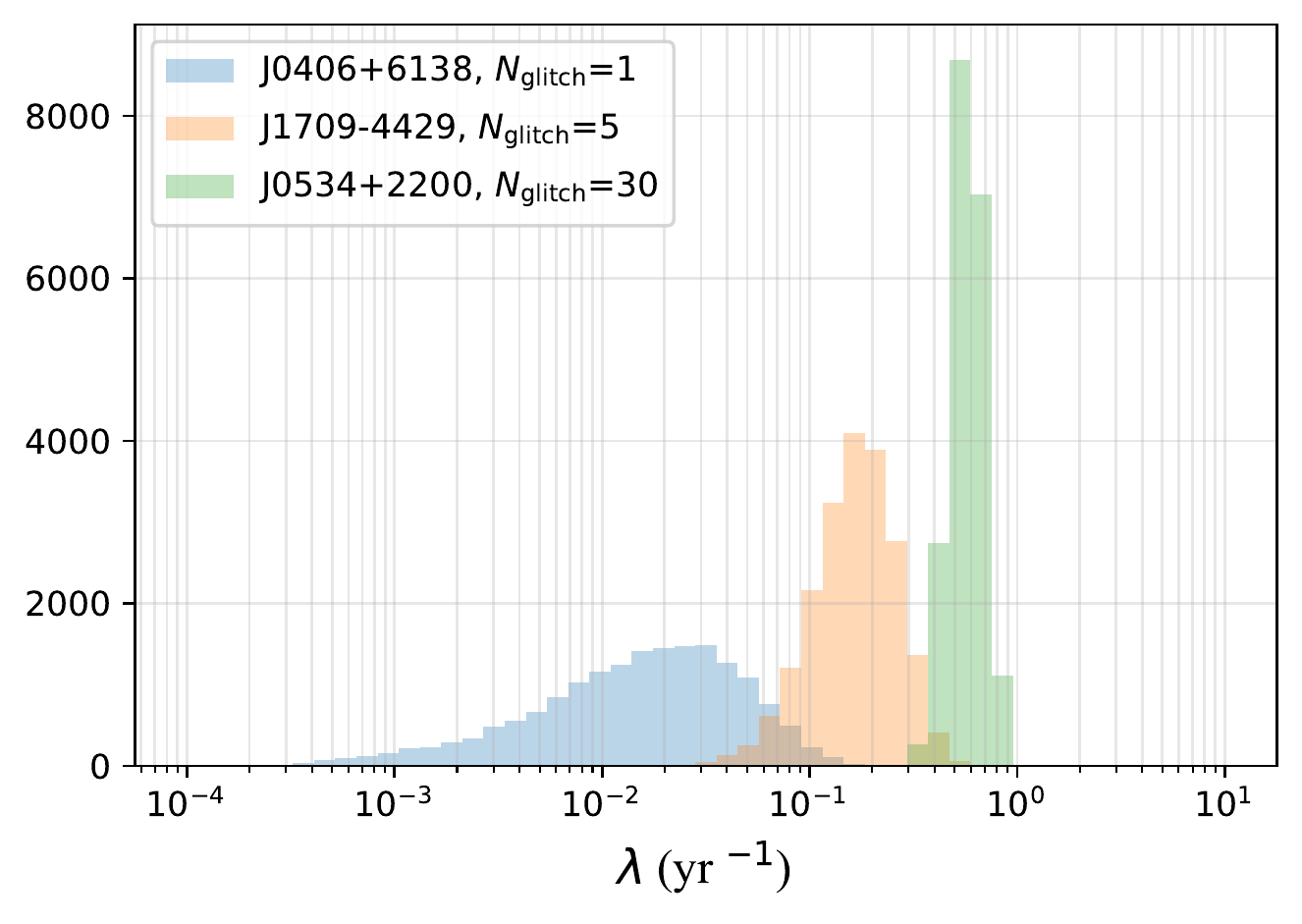}}
  \caption{Example posterior distribution of the glitch rate for three example pulsars with $\Ng=1,5,30$.}
  \label{Fig:lambda_posterior}
\end{figure}

Fig.~\ref{Fig:lambda_posterior} shows the posterior distributions of $\lambda$ for three example pulsars: PSR J0406+6138 with $\Ng=1$, PSR J1709-4429 with $\Ng=5$, and PSR J0534+2200 with $\Ng=30$.  The three examples are chosen arbitrarily to illustrate how the posteriors change with different numbers of glitches, and to demonstrate the unimodality of the posteriors.  As expected the median rate, $\lambda_{\rm med}$, is inversely proportional to $N_{\rm g}$, and the widths of the $90\%$ credible intervals decrease with $\Ng$.

\begin{figure}
  \centering
  \scalebox{0.6}{\includegraphics{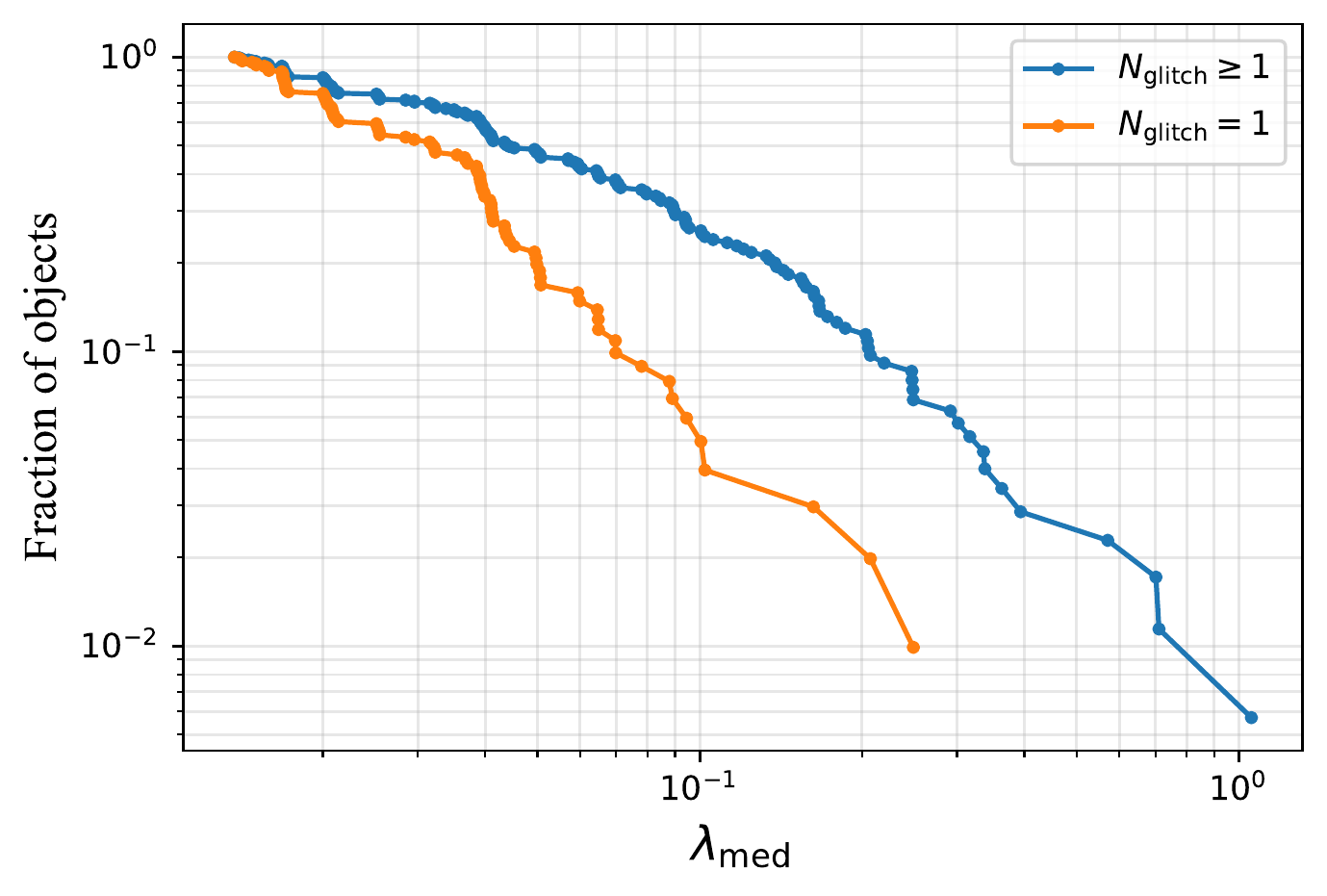}}
  \scalebox{0.6}{\includegraphics{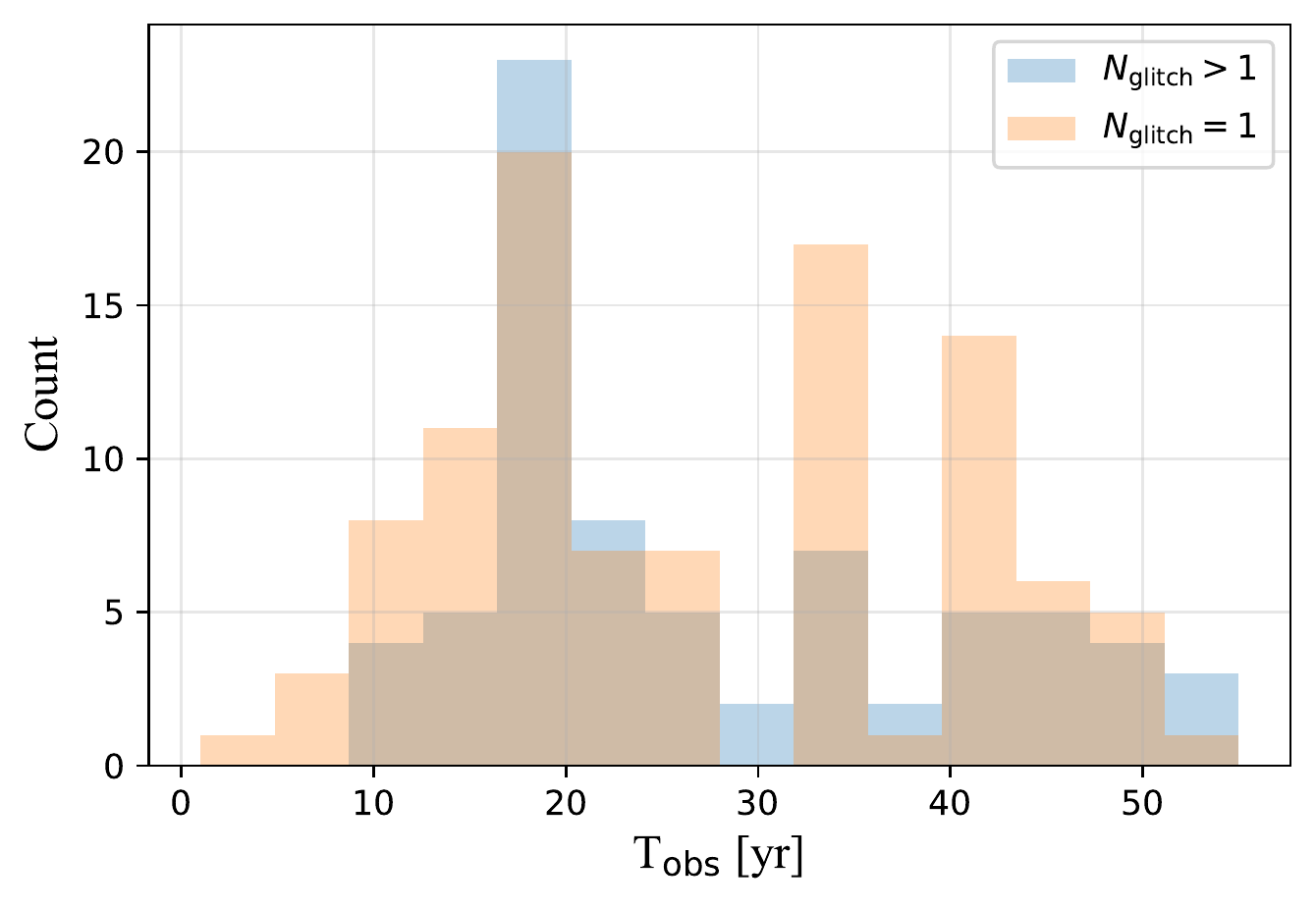}}
  \caption{\emph{Top:} Complementary cumulative distribution of Poisson glitch rate estimates for the glitching pulsars described in Section~\ref{Sec:dataset}. For each point in the plot, the value on the $x$-axis is the estimated $\lambda_{\rm med}$ of one object, and the value on the $y$-axis is the fraction of pulsars with $\lambda_{\rm med}$ equal to or less than that glitch rate.  The distribution of all pulsars with $\Ng\geq1$ is in blue, and the distribution for $\Ng=1$ is in orange.  \emph{Bottom:} Histograms of the observation times for pulsars with $\Ng>1$ (blue) and $\Ng=1$ (orange).
  }
  \label{Fig:rate}
\end{figure}

The median $\lambda_{\rm med}$ and bounds of the $90\%$ credible interval for the \Np{} objects with $\Ng \geq 1$ are given in Table~\ref{Tab:rates}. The distribution of $\lambda_{\rm med}$, is shown in Fig.~\ref{Fig:rate}.  The rate estimates are in the range $0.01\,{\rm yr^{-1}} < \lambda_{\rm med} < 1.05\,{\rm yr^{-1}}$.  The distribution generally looks like a power law $\propto \lambda_{\rm med}^{-1}$ for $\lambda\gtrsim 0.06\,\rm{yr^{-1}}$.
Fig.~\ref{Fig:rate} also shows the distribution of rates for only pulsars with a single recorded glitch.  Interestingly, pulsars with $\Ng=1$ have systematically lower $\lambda_{\rm med}$ per pulsar. The main difference is that the distribution steepens to $\lambda_{\rm med}^{-2}$ for $\lambda \gtrsim 0.06\, \mathrm{yr}^{-1}$. To demonstrate that these low glitch numbers are not simply a result of shorter observation times, we show in the bottom panel of Fig.~\ref{Fig:rate} the distribution of $T_\mathrm{obs}$ for objects with $\Ng=1$ and $\Ng>1$.  The distributions of the two groups are similar visually. Specifically, among pulsars with $\Ng=1$, $44\%$ have been observed for at least 30 years, and $39\%$ of pulsars with $\Ng>1$ have been observed for at least 30 years.   
Excluding pulsars with a low number of glitches from a population analysis is therefore dangerous.

\section{Population-level rate law}\label{Sec:Models}
The previous section infers the present-day instantaneous rate per pulsar, $\lambda$, assuming a homogeneous Poisson process throughout the monitoring interval, $T_\mathrm{obs}$. In this section, we investigate a population model of glitch activity.
We investigate four different phenomenological models of the glitch rate that depend on the frequency $\nu$ and its first time derivative $\nudot$, or the characteristic age\footnote{The definition $\tau = \nu/(2|\dot{\nu}|)$ assumes that the present-day value of $\nu$ is much smaller than the value at birth~\citep{2006puas.book.....L}.} $\tau = \nu/(2 |\dot{\nu}|)$. 
The models are as follows:
\begin{eqnarray}
 \textbf{Model A:} & \lambda_{\rm A} = A\left(\tau/\tau\vref\right)^{-\gamma} \label{Eq:ModelA}\\
 \textbf{Model B:} & \lambda_{\rm B} = A\left(\nu/\nu\vref\right)^{\alpha}\left|\nudot/\nudot\vref\right|^\beta \label{Eq:ModelB}\\
 \textbf{Model C:} & \lambda_{\rm C} = A\left|\nudot/\nudot\vref\right|^\beta \label{Eq:ModelC}\\
 \textbf{Model D:} & \lambda_{\rm D} = A\left(\nu/\nu\vref\right)^{\alpha}. \label{Eq:ModelD}
\end{eqnarray}
These or similar models have previously been considered in the context of timing noise~\citep{2010MNRAS.402.1027H}, and glitch activity~\citep{2017A&A...608A.131F}. 
We use the reference values $\tau\vref=1\ {\rm yr}$, $\nu\vref=1\ {\rm yr^{-1}}$, and $\nudot\vref=1\ {\rm yr}^{-2}$. We note in passing that it is inconvenient to combine $A$ and $\tau_{\rm ref}$ into a single constant, viz. $\lambda = A' \tau^{-\gamma}$ (and similarly for $\nu\vref$ and $\nudot\vref$), because $A'$ has units of (time)$^{\gamma-1}$, making it awkward to compare the normalization of models with different $\gamma$ values in a like-for-like manner. 

The physical motivation for these models is discussed in Section~\ref{Sec:modelMotivation} of this paper and Section 4.1 of~\citet{2017A&A...608A.131F}. In Section~\ref{Sec:methods} we discuss the methods and data sets used in this analysis.  

\subsection{Physical motivation}\label{Sec:modelMotivation}
Several previous surveys have shown evidence that glitch activity is related to observable rotational properties of neutron stars.  Higher glitch rates appear to be correlated with lower $\tau$ and higher $\nudot$
~\citep{2000MNRAS.315..534L,Espinoza2011,2017A&A...608A.131F,2017PhRvD..96f3004A,2019RAA....19...89E}, although there are exceptions. 
A similar conclusion follows from studies of the activity parameter defined by~\cite{McKenna1990}, which find that glitch activity peaks for adolescent pulsars $\sim 10^4\,{\rm yr}$ old~\citep{2000MNRAS.315..534L}. Glitch activity can be related to internal properties of neutron stars, such as their mass, equation of state, and superfluid reservoir~\citep{2020MNRAS.492.4837M,2020MNRAS.tmp.3320W}. 

From the theoretical standpoint, there are physically motivated justifications for Models A--D in Eqs.~\ref{Eq:ModelA}--\ref{Eq:ModelD}.  Many glitch models (see~\cite{2015IJMPD..2430008H} for a recent review) rely on electromagnetic spin down of the neutron star crust to prompt glitches, e.g. via starquakes \citep{2006ApJ...652.1531M,2010MNRAS.407L..54C}, fluid instabilities~\citep{2009PhRvL.102n1101G}, or superfluid vortex avalanches~\citep{2011MNRAS.415.1611W}.  Older neutron stars are then expected to have a lower rate of glitch activity~\citep{2019RAA....19...89E}. In the superfluid vortex avalanche picture, quantum mechanical Gross-Pitaevskii simulations~\citep{2011MNRAS.415.1611W,2015ApJ...807..132M,2019MNRAS.487..702L} and N-body point-vortex simulations ~\citep{2020MNRAS.498..320H} demonstrate that the glitch rate decreases, as the spin-down torque decreases with age; see e.g. Table 9 in~\cite{2011MNRAS.415.1611W} and Table 2 in~\cite{2020MNRAS.498..320H}. The same is true in meta-models, where glitch activity is modeled as a state-dependent Poisson process without specializing to a particular version of the microphysics~\citep{2017MNRAS.470.4307F,2019MNRAS.483.4742C,2019ApJ...885...37M,2020MNRAS.494.3383C}.

The glitch rate may also depend on $\dot{\nu}$ in isolation, as in Model C, without also depending on $\nu$ through $\tau=\nu/(2|\dot{\nu}|)$. Roughly speaking, in Model C, glitches occur when the absolute crust-superfluid lag (or some other stress variable, such as the elastic strain in the crust) exceeds an absolute stress threshold $\Delta\nu_{\rm crit}$, viz. $\lambda^{-1} \dot{\nu} > \Delta\nu_{\rm crit}$. In contrast, in Model A, glitches occur when the fractional crust-superfluid lag exceeds a fractional stress threshold $s_{\rm crit} = (\Delta\nu/\nu)_{\rm crit}$, viz. $\lambda^{-1} \dot{\nu} / \nu > s_{\rm crit}$. Model B is a more general mixture of the physical scenarios underpinning Models A and C. It reduces to Models A, C, and D in the special cases $\alpha=\beta$, $\alpha=0$, and $\beta=0$ respectively. We note that the spin-down torque acts directly on the rigid crust; other components of the star, including possibly a superfluid in the interior, may spin down at different rates over the short and/or long term.

Previous studies have identified $\nudot$ as correlating strongly with glitch activity, where larger glitches dominate this effect~\citep{2000MNRAS.315..534L,Espinoza2011,2017A&A...608A.131F}.  Model C thus serves to test whether a glitch rate that depends only on $\nudot$ better fits the data.

For completeness, in order to fully disentangle the effects of $\nu$ and $\nudot$, we also test a control model that depends only on $\nu$, namely Model D.

\subsection{Bayesian inference}\label{Sec:methods}
In what follows, we compare the relative odds of the four models described in Sec.~\ref{Sec:Models} and produce posterior distributions on the parameters $A,\gamma,\alpha,\beta$.
For each model, we run nested sampling to calculate the Bayesian evidence, and produce posterior distributions.  We again assume a homogeneous Poisson process over the observation time, with the likelihood given by
\begin{equation}
 p\left[\Ng,\Tobs|\lambda_i(\theta_i)\right]= \frac{\left[\lambda_i(\theta_i)\Tobs\right]^{\Ng}e^{-\lambda_i({\theta_i})\Tobs}}{\Ng!}.
  \label{Eq:Likelihood}
\end{equation}
The subscript $i$ indicates the model, and the parameter vector is given by $\theta_i = (A,\gamma), (A,\alpha,\beta), (A,\beta),$ and $(A,\alpha)$ for Models A -- D respectively.
For all models, the prior on $A$ is log uniform between $10^{-5}\ {\rm yr}^{-1}$ and $10^4\ {\rm yr}^{-1}$.  The priors on the exponents in all the models, i.e. $\gamma$, $\alpha$, and $\beta$, are uniform between $-$2.0 and 2.0.  We chose the priors to be relatively uninformative.  For the priors on the exponents, we allow the parameter to be either positive or negative to remain as general as possible.  Validation tests indicate that extending the range of the priors does not affect the posterior distributions reported in Sections~\ref{Sec:PE_glitchers} and \ref{Sec:PE_zeroglitchers}. 

\begin{figure}
  \centering
  \scalebox{0.6}{\includegraphics{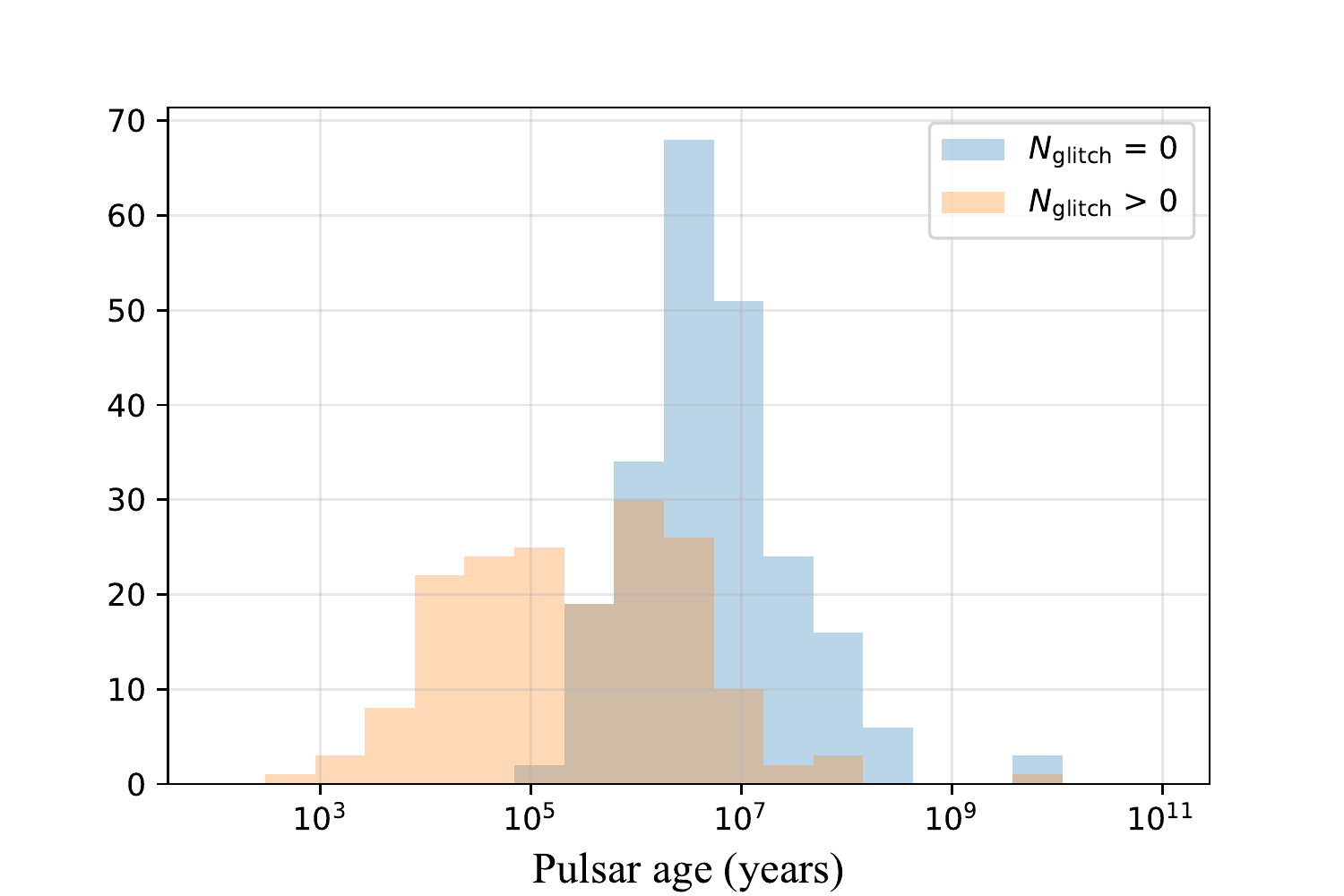}}
  \caption{Histograms of the characteristic ages of the pulsars used in this analysis for pulsars with recorded glitches (blue), and no recorded glitches (orange). Pulsars with no recorded glitches tend to have higher characteristic ages.}
  \label{Fig:age_histogram}
\end{figure}

We run each model using two different data sets.
In the first data set there are \Np{} glitching pulsars, with rates between $1.98 \times 10^{-2}\, {\rm yr^{-1}} < \lambda < 3.47\, {\rm yr^{-1}}$. The ages are in the range $728\, {\rm yr} < \tau < 5.06 \times 10^{9}\, {\rm yr}$, and the observation times are in the range $3\, {\rm yr} < \Tobs < 53\, {\rm yr}$.
However, many pulsars have no recorded glitches at all, despite being monitored for a substantial length of time.  If $\lambda$ is inferred to depend on $\tau$, $\nu$, or $\nudot$, pulsars with no observed glitches provide extra evidence to strengthen or weaken that inference.
As described in Section~\ref{Sec:dataset}, we include a set that have been routinely monitored by UTMOST and have no recorded glitches.  There are a total of \Nz{} pulsars in this set.  The characteristic ages are in the range $1.61 \times 10^5\, {\rm yr} \leq \tau \leq 7.58 \times 10^9\, \mathrm{yr} $.  
The distributions of characteristic ages for pulsars with $\Ng\geq1$ and $\Ng=0$ are shown in Fig.~\ref{Fig:age_histogram}.  We see that pulsars with $\Ng=0$ are older than those with $\Ng\geq1$, with $\langle \tau \rangle = 1.0 \times 10^8\,{\rm yr}$ and $ 3.1\times10^{7}\,{\rm yr}$ for the two populations respectively.

\section{Glitching pulsars with $\Ng\geq 1$}\label{Sec:GlitchResults}

\subsection{Model selection}\label{Sec:BayesFactors}
The Bayes factor $\mathcal{B}$, calculated as the ratio of the Bayesian evidences of two models, quantifies the odds of one model relative to the other, assuming equal prior odds.
The log Bayes factors relating Models A--D, inferred from the data set containing $N_g\geq1$, are presented in the top half of Table~\ref{Table:BFs_glitchers}. Model A is preferred over all other models.  Model B has the second highest evidence, with $\log\mathcal{B}_{\rm AB} = \logBABglitch$.  Model D, which depends only on the pulsar frequency $\nu$, is strongly disfavored.

The preference for Model A arguably points towards a glitch trigger threshold involving the fractional stress (e.g. crust-superfluid angular velocity lag) instead of the absolute stress, as discussed in Section~\ref{Sec:modelMotivation}, although it must be emphasized that many other scenarios are possible too. The preference for Model A is also consistent with previous studies involving glitches of all sizes~\citep{2019RAA....19...89E}.

\subsection{Parameter estimation}\label{Sec:PE_glitchers}
In addition to calculating the evidence, we also produce posterior distributions on the parameters of all four models.  The medians and 90\% credible interval bounds are summarized in Table~\ref{Table:PE}. The full posterior distributions for all models are plotted for completeness in Appendix~\ref{Ap:Posteriors}.    

For Model A, the preferred model, the posterior on $\gamma$ shows zero support for $\gamma\leq0$, indicating that the glitch rate decreases with age.  The posterior distributions on $\alpha$ and $\beta$ for Model B show zero support $\beta\leq0$ or $\alpha\geq0$, consistent with a lower glitch rate for older pulsars. Model C also shows zero support for $\beta\leq0$. Interestingly Model D, which which depends only on $\nu$, predicts $\alpha>0$ but it is disfavored strongly, as shown in Table~\ref{Table:BFs_glitchers}.

Of particular interest are the posterior distributions for Models A and B.  The medians of the posteriors estimate $|\alpha|\approx\beta\approx\gamma$, with $\gamma=\ModelAgammaglitch$, $\alpha=\ModelBalphaglitch$, $\beta=\ModelBbetaglitch$. Fig.~\ref{Fig:GammaBetaAlpha} shows the posterior distributions for $|\alpha|$, $\beta$, and $\gamma$ plotted together.  We see the posteriors peak at the same value.  The widths of the posteriors for $\gamma$ and $\beta$ are similar, though the posteriors are about twice as wide for $|\alpha|$. All in all, the tendency of Model B (in which $\alpha$ and $\beta$ are unconstrained) to return $|\alpha| \approx \beta$ and hence $\lambda \propto \tau$, as in Model A, engenders extra confidence in the conclusion that Model A is preferred for the data set analyzed here. There is also statistical flexibility to accommodate the scaling $\lambda \propto \dot{\nu}$ discovered by~\citet{2017A&A...608A.131F}, because the latter scaling is dominated by glitches above a certain size, whereas Table ~\ref{Table:BFs_glitchers} analyses small and large glitches on an equal footing. This issue is discussed further in Sec.~\ref{Sec:conclusion}.

\section{Glitching and nonglitching pulsars}\label{Sec:NonglitchResults}
\subsection{Model selection}
The Bayes factors for the analysis that includes $\Ng=0$ pulsars are shown in the bottom half of Table~\ref{Table:BFs_glitchers}. We see that the Bayes factors change from Sec.~\ref{Sec:BayesFactors}, but the overall conclusions remain the same.  Model A is still preferred over all other models, and Model D is disfavored. 

In general, the magnitudes of the Bayes factors increase when including $\Ng=0$ pulsars. The exception is $\log\mathcal{B}_{\rm AB}$ which decreases slightly from $\logBABglitch$ to $\logBABall$.  However the before-and-after values are still consistent within one standard deviation.

We emphasize again that an unknown fraction of the pulsars with $\Ng=0$ may turn out to glitch quasiperiodically, as flagged in Sec.~\ref{Sec:DataGlitchers}. The analysis should be redone in the future, once more data become available and the situation is clearer, with Eq.~\ref{Eq:Poisson} replaced by a likelihood suitable for quasiperiodic activity in quasiperiodic objects~\citep{2019MNRAS.483.4742C}.

\begin{table*}
\begin{center}
\begin{tabular}{c c c c c }
\hline
$\Ng\geq1$  & Model A & Model B & Model C & Model D \\
  \hline
Model A  & - & $\logBABglitch$ &  $\logBACglitch$ & $\logBADglitch$ \\
Model B & $-\logBABglitch$ & - & $\logBBCglitch$ & $\logBBDglitch$ \\
Model C & $-\logBACglitch$ & $-\logBBCglitch$ & - & $\logBCDglitch$ \\
Model D  & $-\logBADglitch$ & $-\logBBDglitch$ & $-\logBCDglitch$ & - \\
\hline
$\Ng\geq0$   & Model A & Model B & Model C & Model D \\
  \hline
 Model A  & - & $\logBABall$ &  $\logBACall$ & $\logBADall$ \\
Model B & $-\logBABall$ & - & $\logBBCall$ & $\logBBDall$ \\
Model C & $-\logBACall$ & $-\logBBCall$ & - & $\logBCDall$ \\
Model D  & $-\logBADall$ & $-\logBBDall$ & $-\logBCDall$ & - \\
\hline
\end{tabular}
\end{center}
 \caption{Log Bayes factors between the rate laws given by Models A--D in Equations~\ref{Eq:ModelA}--\ref{Eq:ModelD}, for the data containing $\Ng\geq1$ (top) and $\Ng\geq0$ (bottom).  Bayes factors are of the form $\log\mathcal{B}_{\rm row, column}$.}
\label{Table:BFs_glitchers}
\end{table*}%

\begin{table*}
\begin{tabular}{c c c c c}
\hline
$\Ng\geq1$   & A & $\gamma$ & $\alpha$ & $\beta$ \\
  \hline
Model A  & $\ModelAAglitch$ & $\ModelAgammaglitch$ &  - & - \\
Model B & $\ModelBAglitch$ & - & $\ModelBalphaglitch$ & $\ModelBbetaglitch$ \\
Model C & $\ModelCAglitch$ & - & - & $\ModelCbetaglitch$ \\
Model D  & $\ModelDAglitch$ & - & $ \ModelDalphaglitch$ & - \\
\hline
$\Ng\geq0$   & A & $\gamma$ & $\alpha$ & $\beta$ \\
  \hline
Model A  & $\ModelAAall$ & $\ModelAgammaall$ &  - & - \\
Model B & $\ModelBAall$ & - & $\ModelBalphaall$ & $\ModelBbetaall$ \\
Model C & $\ModelCAall$ & - & - & $\ModelCbetaall$ \\
Model D  & $\ModelDAall$ & - & $ \ModelDalphaall$ & - \\
\hline
\end{tabular}
 \caption{Parameter estimates for Models A--D in Equations~\ref{Eq:ModelA}--\ref{Eq:ModelD}.  \emph{Top:} Median and 90\% credible intervals when including pulsars with $\Ng\geq1$. \emph{Bottom:} Median and 90\% credible intervals analyzing pulsars with $\Ng\geq0$.}
\label{Table:PE}
\end{table*}%

\begin{figure}
\centering
  \scalebox{0.5}{\includegraphics{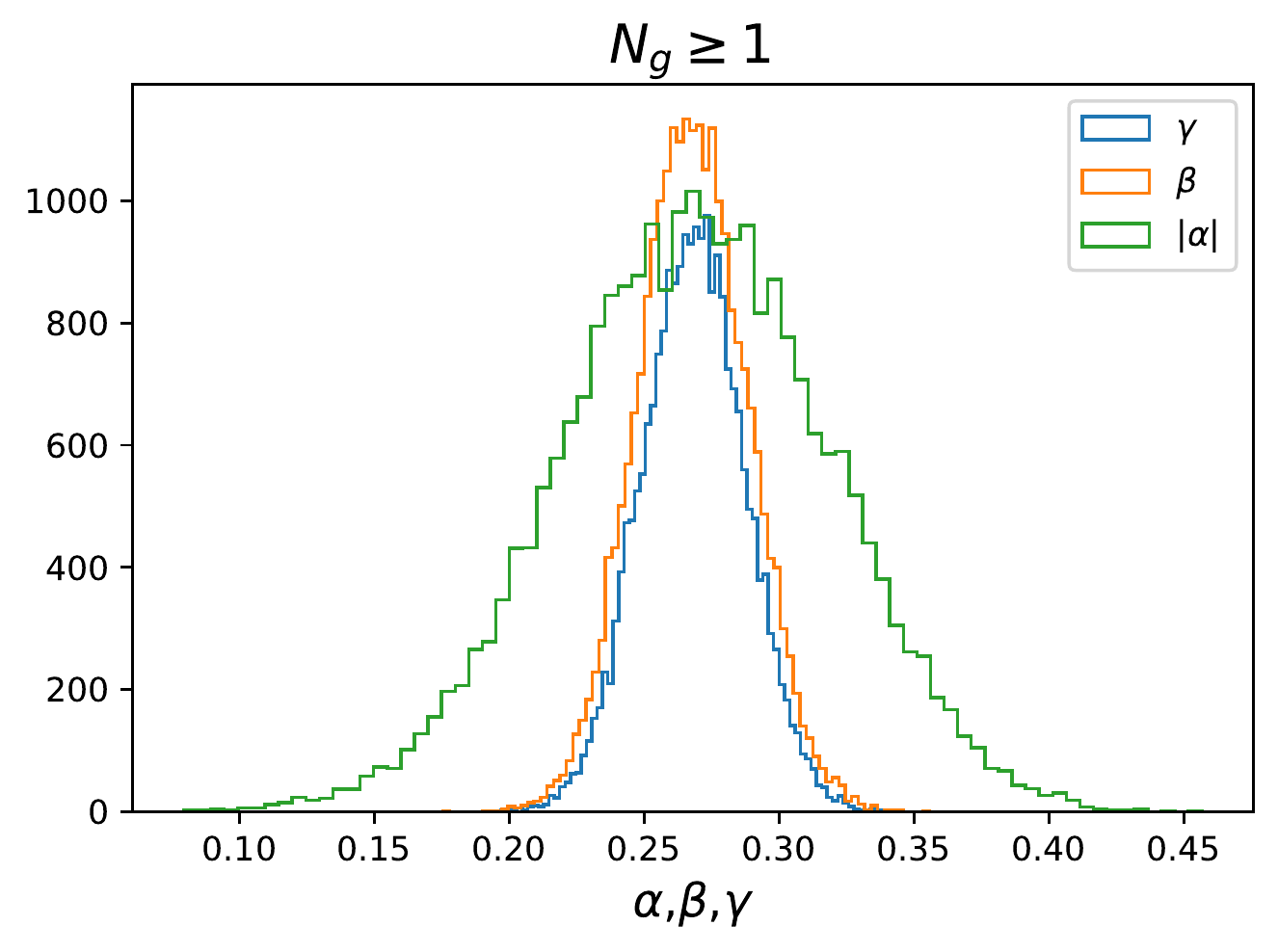}}
  \scalebox{0.5}{\includegraphics{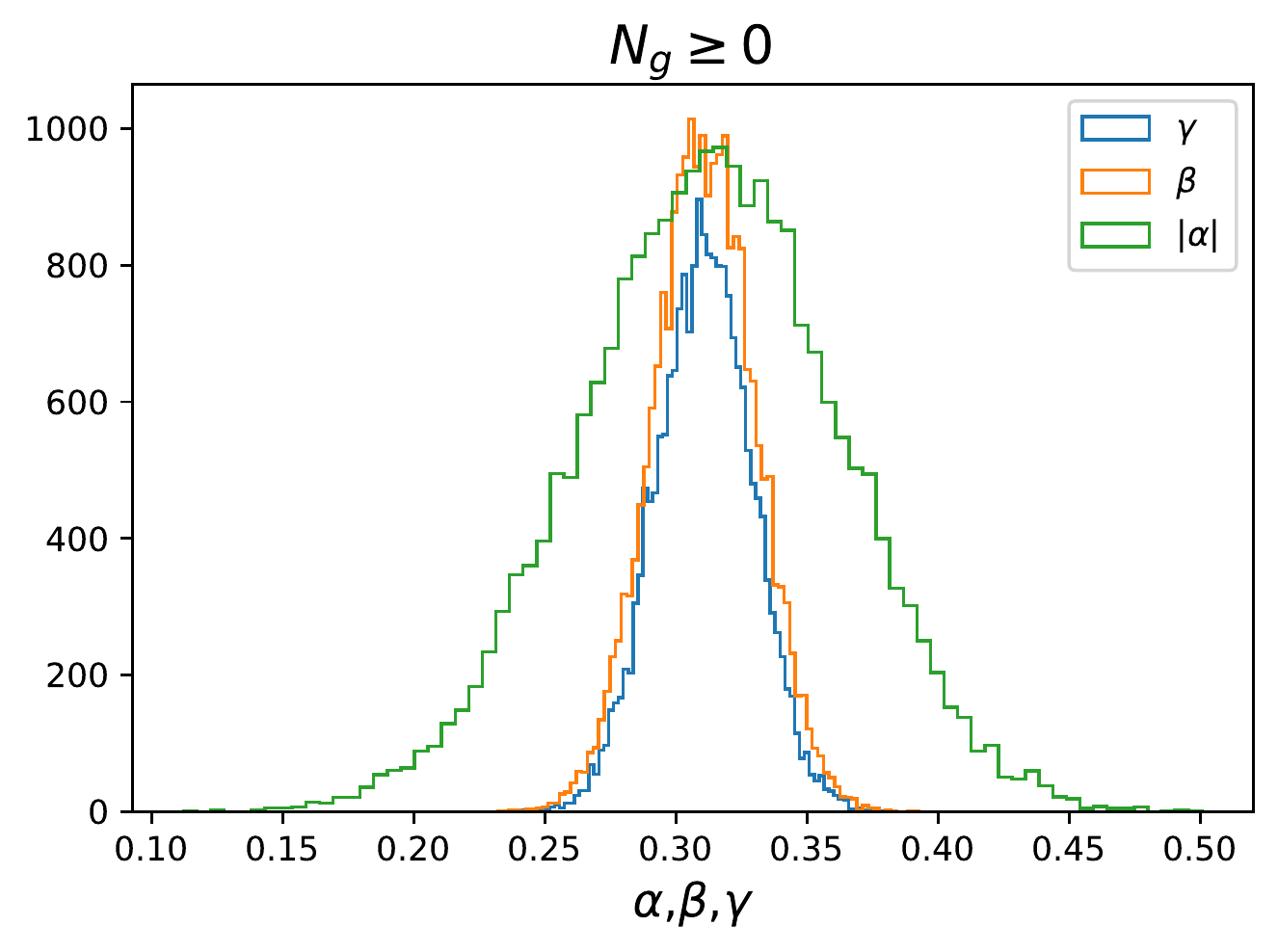}}
  \caption{Posterior distributions for $\gamma$ from Model A, and $\alpha$ and $\beta$ from Model B. \emph{Top:} Data comprising $N_g\geq1$ pulsars. \emph{Bottom:} Data comprising $N_g\geq0$ pulsars.}
  \label{Fig:GammaBetaAlpha}
\end{figure}

\subsection{Parameter estimation}\label{Sec:PE_zeroglitchers}
For the $\Ng \geq 0$ data set, the median and 90\% credible intervals for the parameters are given in the bottom of Table~\ref{Table:PE}. The full posterior distributions are plotted for completeness in Appendix~\ref{Ap:Posteriors}.  For all models, including the $\Ng=0$ pulsars shifts the posteriors relative to the $\Ng \geq 1$ data set. Indeed including data from non-glitching pulsars prompts one to infer that the pulsar glitch rate is more sensitive to pulsar age or spin-down rate, i.e. a given change in pulsar age leads to a larger change in the glitch rate.

Though the values change, the overall trends of the posteriors are consistent with the results discussed in Sec.~\ref{Sec:PE_glitchers}.  In Model A, there is still zero posterior support for $\gamma\leq0$. In Model B there is still zero support $\beta\leq0$ or $\alpha\geq0$ in Model B.  Focusing on Models A and B, we see the same interesting results $|\alpha|\approx\beta\approx\gamma$, with $\gamma=\ModelAgammaall$, $\alpha=\ModelBalphaall$, $\beta=\ModelBbetaall$.  The posteriors for $|\alpha|$, $\beta$, and $\gamma$ are shown in Fig.~\ref{Fig:GammaBetaAlpha}.  The posteriors for $\gamma$ and $\beta$ have similar maximum values and widths, while the posterior on $|\alpha|$ is about twice as wide as the other two. As stated in Sec.~\ref{Sec:PE_glitchers}, the unconstrained finding $|\alpha|\approx\beta\approx\gamma$ further supports the conclusion that Model A is preferred by the data.

\section{Biases}\label{Sec:Biases}
There are at least two observational biases that impact the parameter estimation in Sections~\ref{Sec:GlitchResults} and ~\ref{Sec:NonglitchResults}. First, $\Tobs$ may be incorrectly estimated. Second, the glitch catalog could be incomplete, especially for small glitches~\citep{Espinoza2011,2019A&A...630A.115F,2021MNRAS.tmp.2433L}.  Here we discuss these limitations.

\subsection{Observation time}\label{Sec:TobsBiases}
It is challenging to retrieve complete records of $T_{\rm obs}$ from the public literature for some pulsars without consulting hard-to-find documents such as telescope logs. In this section, 
we follow the lead of~\cite{2019A&A...630A.115F} and run Monte Carlo tests to gauge the impact of overestimating $\Tobs$. We focus on Model A, which is preferred by the data as shown in Sections~\ref{Sec:GlitchResults} and~\ref{Sec:NonglitchResults}.

In previous sections, for pulsars with $\Ng\geq1$ we define the start of the observation interval as the date of the discovery publication.  Many pulsars however are not continuously monitored after their discovery, and long enough observational gaps could potentially lead to the missed detection of a glitch. Therefore $\Tobs$ may be overestimated if the discovery date is used.

To investigate the impact of overestimating $\Tobs$, we simulate pulsar observations and compare the posterior distributions recovered when we artificially bias $\Tobs$. The procedure for simulating pulsar observations is as follows.  We draw a set of $\tau$ values from the distribution of ages of the pulsars listed in Sec.~\ref{Sec:dataset}.  We draw $A$ and $\gamma$ from the priors described in Sec.~\ref{Sec:methods}.  We then calculate the simulated glitch rate $\lambda$ per pulsar from Model A via Eq.~\ref{Eq:ModelA}. We select $\Tobs$ for each object by drawing from a uniform distribution between 10 and 60 years.  Once we have $\lambda$ and $\Tobs$ for each pulsar, we draw an observed number of glitches from the Poisson distribution, as given in Eq.~\ref{Eq:Poisson}. We then artificially bias $\Tobs$ and produce posterior distributions on $A$ and $\gamma$.

Let $\Tobs'$ denote the overestimated observation times. First, we look at the extreme case $\Tobs'=2\Tobs$. The choice $\Tobs'=2\Tobs$ is pessimistic, but we include it for illustrative purposes.  When running over 100 data realisations, we see that this extreme overestimation of $\Tobs$ leads to systematic underestimations of both $A$ and $\gamma$.  The median of the posterior on $A$ tends to be 60\% lower than the true value, and the true value is never contained within the 90\% credible intervals.  The median of the posteriors on $\gamma$ are around 5\% lower than the injected values, and the injected value is inside the 90\% credible interval 40\% of the time.

\begin{figure}
  \scalebox{0.6}{\includegraphics{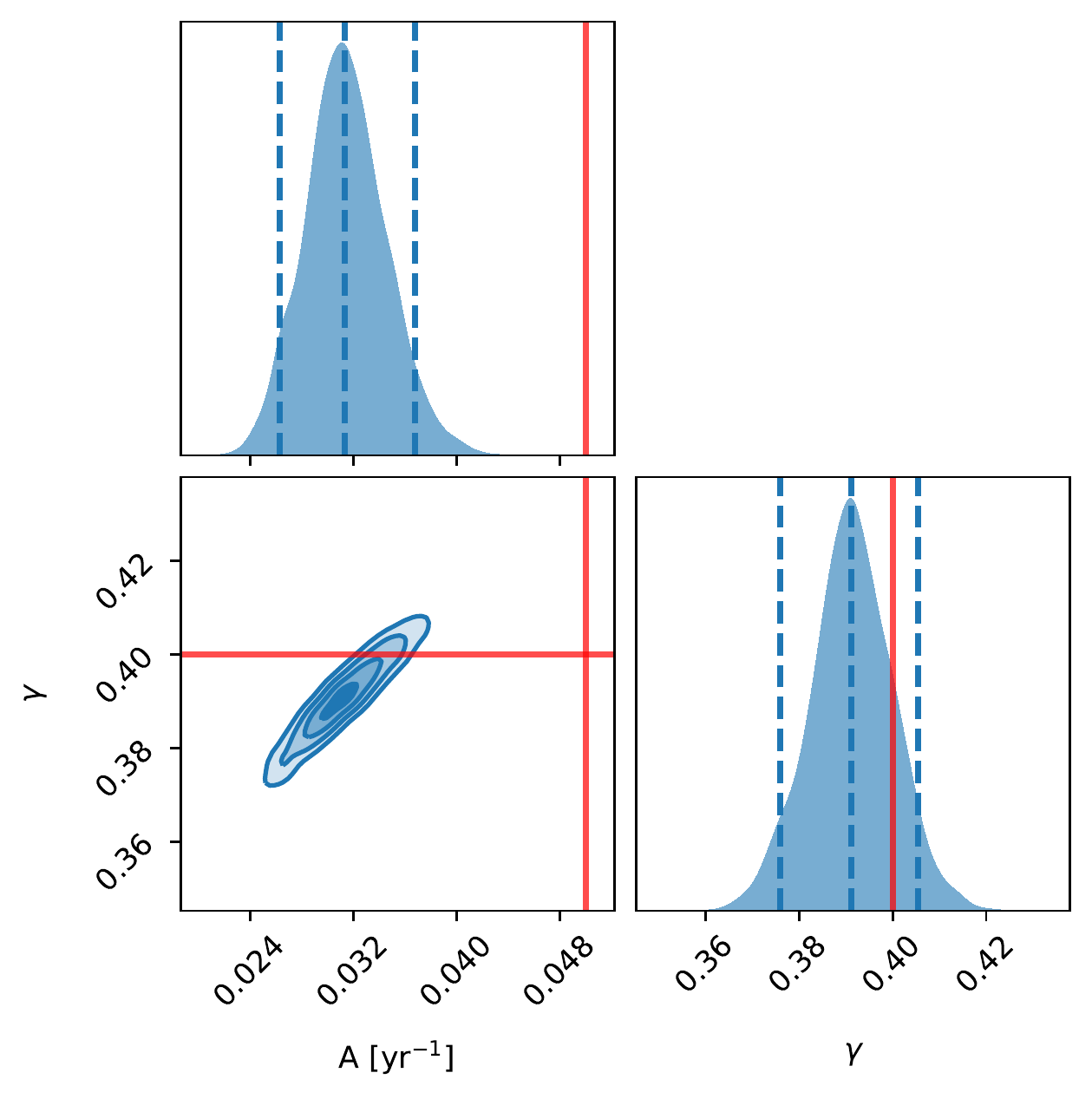}}
  \caption{Posterior distributions for the age-rate law from simulated pulsar observations with $\Tobs$ overestimated by $\Tobs'=u\cdot\Tobs$, with $u\in[1.0,2.0]$ as described in Section~\ref{Sec:TobsBiases}. The red lines mark the values of $A$ and $\gamma$ used to produce the data.}
  \label{Fig:TobsOverEst}
\end{figure}

Next we look at a more realistic case with $\Tobs'= u\Tobs$ where $u$ is distributed randomly and uniformly in the range $1.0 \leq u \leq 2.0$.  We again see a tendency for $A$ and $\gamma$ to be underestimated. An example of one data realisation is shown in Fig.~\ref{Fig:TobsOverEst}. We find that the median of the posterior on $A$ tends to underestimate the true value by about 40\% (i.e. less than for $T'_{\rm obs}=2T_{\rm obs}$) though the true value of $A$ lies in the 90\% credible interval just 1\% of the time. The posterior distributions on $\gamma$ however show less biasing than we see for $A$, with the median value underestimating the true value by less than 5\%, and the 90\% credible intervals encompassing the true value 70\% of the time.

Finally we look at the case where $\Tobs$ is overestimated by a constant and arbitrary (yet reasonable) amount, viz. $\Tobs'=\Tobs+2\ {\rm yr}$. Here we see minimal impact on the posterior distributions.  The 90\% credible intervals contain the true value of $A$ 85\% of the time perhaps indicating a slight bias, and the true value of $\gamma$ is enclosed in the 90\% credible interval 90\% of the time as expected.

If $\Tobs$ is systematically overestimated in the analysis presented in this paper, then the posterior distributions on $A$ and $\gamma$ may underestimate the true values. We note however that the simulations in this study indicate that these biases may affect $A$ more than $\gamma$.

\begin{figure}
  \centering
  \scalebox{0.5}{\includegraphics{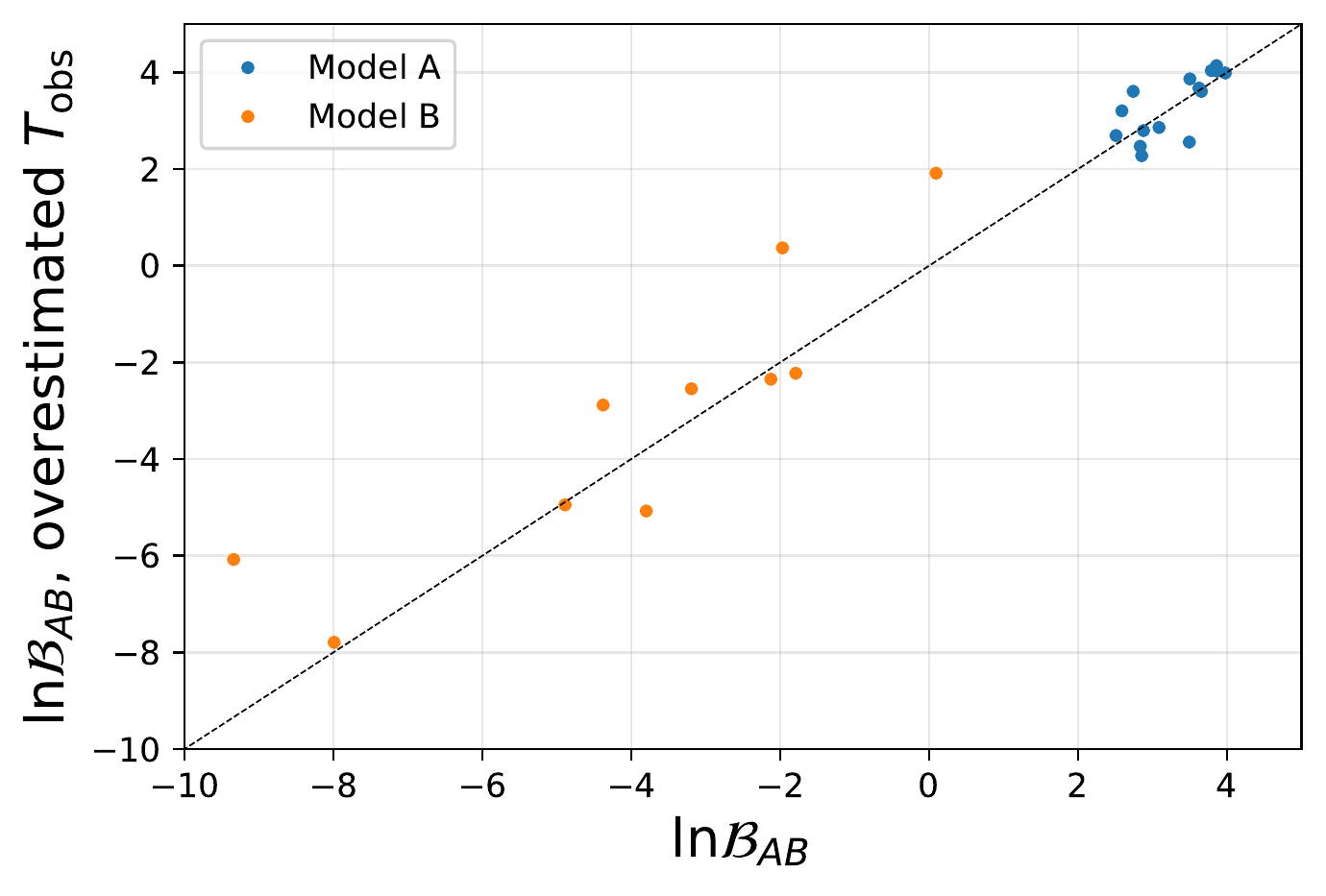}}
  \caption{Comparison of the log of the Bayes factors between Models A and B for simulated data and an artificially overestimated $\Tobs$.  Blue and orange dots indicate simulated data produced using the glitch rate from Models A and B respectively. The dashed black line is the diagonal, where the abscissa equals the ordinate.}
  \label{Fig:BF_biasing}
\end{figure}

As a further safety check, we investigate the effect of overestimating $\Tobs$ on Bayes factors. We find no systematic bias. Fig.~\ref{Fig:BF_biasing} compares the log Bayes factors between Models A and B for simulated data, both with and without artificially overestimating $\Tobs'=u\Tobs$ with $u\in[1.0,2.0]$.  The points generally scatter about the diagonal. In addition to simulating data according to Model A (blue points), we also simulate data using the rate law from Model B (orange points), using the same procedure described above, drawing $\alpha$ and $\beta$ from prior distributions instead of $\gamma$. Even when $\Tobs$ is overestimated, we see that the Bayes factors prefer the correct model.

\subsection{Completeness of glitch catalogues}\label{Sec:completeness}
Incompleteness of glitch catalogs affects glitch rate-age studies in two ways. First, it compromises the $\lambda$ estimate for individual objects, by miscounting the number of glitches during $T_{\rm obs}$. Second, it compromises the $A$ and $\gamma$ estimates at the population level, if the observational biases causing incompleteness are systematically different in younger and older pulsars. 

It is unlikely that any glitch catalogue is complete.
Many pulsars are observed infrequently, so a unique phase-connected timing solution cannot always be constructed between ToAs.
If a glitch occurs inside a sufficiently lengthy ToA gap, its epoch cannot be determined uniquely; it may also be challenging to distinguish between a single, large glitch and multiple, smaller glitches
\citep{Janssen2006,Yu2017,2020ApJ...896...78M}.
Another issue is the detectability of small glitches, which may not be resolvable above the timing noise, or, if they occur shortly after a large glitch, may be absorbed into the fitted parameters of the larger glitch's recovery
\citep{Wong2001,Espinoza2014}. Historically it has been hard to quantify these biases systematically, although recent advances in Bayesian parameter estimation, e.g. with {\sc temponest}~\citep{10.1093/mnras/stt2122,Lower_2020}, and hidden Markov models~\citep{2020ApJ...896...78M,2021MNRAS.tmp.2433L}, offer promising pathways for future work.

Uncertainty also exists in the classification of small glitches. 
Some authors distinguish between regular (or `macro') glitches and `micro-glitches'
(e.g. \citet{Cordes1985,1995MNRAS.277.1033D,Chukwude2010,Onuchukwu2016}).
However, it is not clear that macro-glitches are always accompanied by increases in the spin-down rate
\citep{Espinoza2011}.
It is also unclear if the distinction between these two classes of events is strictly adhered to in the literature.
As an example,
\citet{Janssen2006} found four small 
$(10^{-11} \lesssim \Delta \nu / \nu \lesssim 10^{-10})$ glitches in PSR J0358+5413 in six years of well-sampled observing data, which are widely viewed as macro glitches despite their small sizes.
These glitches are recorded in the ATNF pulsar glitch catalogue
\citep{2005AJ....129.1993M}, and in 
\citet{Espinoza2011},
but not the online Jodrell Bank glitch catalogue\footnote{Based on the data from 
\citet{Janssen2006},
one expects $\approx$ 10 more glitches to have been detected in this pulsar, however, at the time of writing, no new glitches have been reported since April 2004.}. 

Several authors have examined the issue of completeness in individual pulsars or in particular datasets.
\citet{Janssen2006}, 
in their analysis of PSR J0358+5413, performed Monte Carlo simulations to test their glitch detection method, and concluded that they were able to detect glitches down to a threshold of $\Delta \nu / \nu \sim 10^{-11}$, about the size of the smallest glitch they detected in that pulsar.
\citet{Janssen2006}
suggested that the inconsistency with a power-law PDF at small $\Delta \nu$ is due to the difficulty of detecting glitches with $\Delta \nu / \nu \lesssim 10^{-9}$, rather than a physical mechanism restricting the minimum glitch size. 
\citet{Espinoza2014} 
performed an analysis of 29 years of timing data for the Crab pulsar (PSR J0534+2200).
They selected events with $\Delta \nu > 0$ and $\Delta \dot{\nu} < 0$ between consecutive ToAs, identifying all previously detected glitches, as well as several hundred candidate events with smaller $\Delta \nu$ than the previously detected glitches.
They found a gap of around half a decade in $\Delta \nu$ between the largest of the glitch candidates and the smallest previously detected glitch.
A similar analysis for glitches in the Vela pulsar, however, finds no such evidence of a minimum glitch size
\citep{Espinoza2020}.

\citet{Yu2017} 
performed a statistical analysis on the completeness of the data set from 
\citet{Yu2013},
which discovered 107 glitches in 36 pulsars from $\sim 20$ yr of timing data from the Parkes radio telescope.
\citet{Yu2017} performed Monte Carlo simulations to construct synthetic TOAs for each pulsar, added glitches and searched the synthetic data for glitches using a nested sampling method to determine the glitch parameters. 
It is found that the expected and observed PDFs match except for below $\Delta \nu / \nu \lesssim 10^{-7}$, where glitches are harder to detect
\citep{Yu2013}.

\begin{figure}
\includegraphics[scale=0.4]{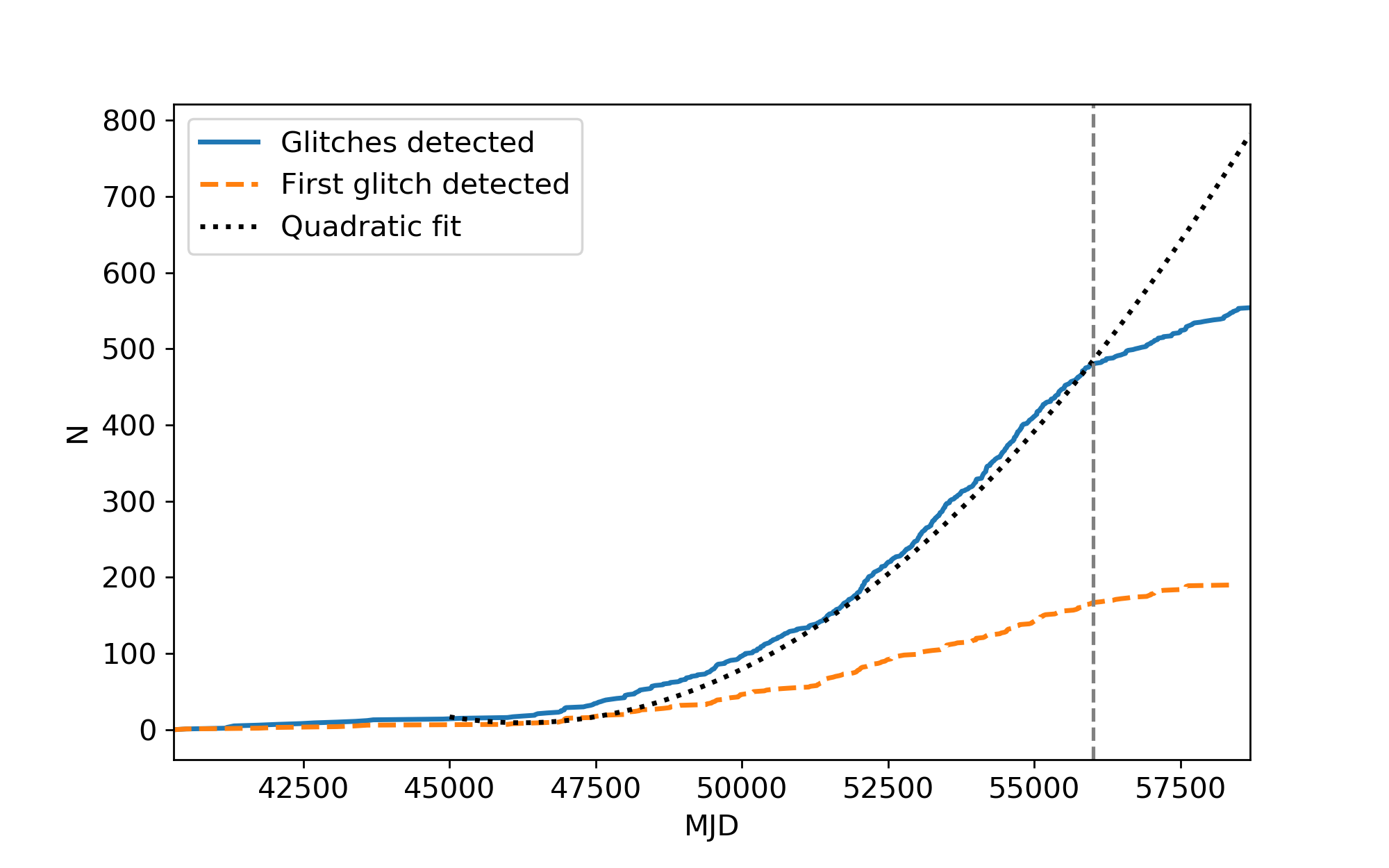}
\caption{Cumulative detection rates of pulsar glitches (blue curve) and glitching pulsar discovery (orange dashed curve) versus time.
Dotted black curve is a quadratic fit to the blue curve for $45000 < \mathrm{MJD} < 56000$.
Data taken from the Jodrell Bank glitch catalogue \url{http://www.jb.man.ac.uk/pulsar/glitches.html}
\citep{Espinoza2011}.}
\label{fig_chap1:discovery history}
\end{figure}

It may be possible to identify incompleteness in glitch catalogues by comparing the rate of discovery of glitches and glitching pulsars.
Two things are expected in such a comparison if the glitch catalogue is complete.
Firstly, the rate at which glitches are detected should increase as more glitching pulsars are discovered, as previously discovered objects continue to glitch.
Secondly, the rate at which glitches are detected should exceed the rate at which new glitching pulsars are discovered, for the same reason.
Because glitches are a stochastic process, both effects are only expected to be observable on time scales longer than the average time between observations for the most frequently glitching pulsars.
In Figure \ref{fig_chap1:discovery history}, we show the cumulative number of detected pulsar glitches and also the cumulative number of glitching pulsars discovered as a function of time over the past 50 years.
The data are taken from the discovery epochs recorded on the Jodrell Bank glitch catalogue 
\citep{Espinoza2011}.
We take the date of discovery to be the epoch of the first detected glitch, because many non-glitching pulsars are not regularly monitored, so there may be an artificial lag between the discovery of a pulsar and the first glitch detection in that pulsar.

Figure \ref{fig_chap1:discovery history} shows that the rate of glitch detection exceeds the rate of glitching pulsar discovery throughout the history of pulsar astronomy.
In the past 10 years, however, the glitch detection rate has decreased, from $\approx 20$ glitches yr$^{-1}$ during 50000 $\lesssim$ MJD $\lesssim$ 56000 to $\approx 10$ yr$^{-1}$ during $56000 \lesssim$ MJD 
$\lesssim 59000$.
If the rate of glitching pulsar discovery is approximately linear in time then the rate of glitch detection should be quadratic in time, assuming that the glitch rate is roughly the same for all pulsars.
We include a quadratic fit for the glitch detection rate in Figure \ref{fig_chap1:discovery history}, based on the data from $45000 <$ MJD $< 56000$.
This suggests that $\sim 200$ glitches may have occurred in known glitching pulsars in the last 10 years that have not been recorded in glitch catalogues. One possible contributing factor is that the Jodrell Bank glitch catalog was not regularly updated with unpublished glitches detected by the Parkes radio telescope since the publication of~\cite{Yu2013}.  Recently, ~\cite{2021RAA....21..154L} reported 30 glitches in data from the Parkes telescope, 26 of which had not been published previously. However we note that these 26 glitches explain only $~10\%$ per cent of the shortfall of $\approx 200$ glitches at MJD 59000 in Fig.~\ref{fig_chap1:discovery history}.

Fortunately, current and upcoming radio instruments, such as UTMOST, MeerKAT, SKA, and CHIME will be able to time many pulsars much more frequently and with greater sensitivity, so that in coming years many more glitches will be detected with greater completeness
\citep{Bailes2018,Levin2018,Jankowski2019}.
As well as improved instruments, improved glitch-finding algorithms are being developed that will ensure consistent classification of timing irregularities in pulsars
\citep{2020ApJ...896...78M,Lower_2020,2021MNRAS.tmp.2433L,2021ApJS..255....5C}.

\section{Discussion}\label{Sec:conclusion}

In this paper we use \Np{} pulsars with at least one recorded glitch, and \Nz{} pulsars with zero recorded glitches to perform a Bayesian analysis of four phenomenological glitch rate laws which depend on $\tau$, $\nu$, and/or $\nudot$.

We assume first that glitch activity of any given pulsar is a homogeneous Poisson process with rate $\lambda$.  
We therefore exclude from the analysis three objects that are known to glitch quasiperiodically (PSR J0835-4510, PSR J0537-6910, PSR J1341-6220), noting also that other objects retained in the sample may turn out to glitch quasiperiodically, when more data become available in the future.
We estimate $\lambda$ for all pulsars with $\Ng\geq1$ and find that the median rate satisfies $0.01\,{\rm yr^{-1}} < \lambda_{\rm med} < 1.05\,{\rm yr^{-1}}$.
We also find that objects with $\Ng=1$ have lower $\lambda_{\rm med}$ than objects with $\Ng \geq 1$, which is not a selection effect of short observing times.
We then use nested sampling to calculate the Bayesian evidence for the four rate laws, and find that Model A, $\lambda=A\left(\tau/\tau_{\rm ref}\right)^{-\gamma}$, is preferred with  $A=\ModelAAglitch\ \rm{yr}^{-1}$, and $\gamma=\ModelAgammaglitch$.  More than 99.99\% of the probability weight is found at $\gamma>0$.  This indicates that the glitch rate does indeed decrease with characteristic age. 
Model B, of the form $\lambda = A\left(\nu/\nu\vref\right)^{\alpha}\left|\nudot/\nudot\vref\right|^\beta$, has the next largest evidence after Model A with $\log\mathcal{B}_{AB}=\logBABglitch$, and interestingly the posteriors show $|\alpha|\approx\beta\approx\gamma$. The Bayes factors and posterior distributions both support the hypothesis that glitch activity is related to the characteristic spin-down age of the pulsar.

We then repeat the analysis including a subset of UTMOST-monitored pulsars with no known associated glitches. The preferred model is still the model that depends on the spin-down age $\tau$. The posterior distributions on $A$ and $\gamma$ shift to $A=\ModelAAall\ \rm{yr}^{-1}$, and $\gamma=\ModelAgammaall$. These results still support a rate-age relation, but the parameters shift because the objects with $\Ng= 0$ are numerous and contain valuable information.

One physically plausible relation between rate and age is that the two quantities are inversely proportional, i.e. $\gamma=1$.  
This has been proposed in the literature, for example as an explanation for the low but nonzero glitch activity in millisecond pulsars; see e.g. Section 4.2 in ~\cite{2015ApJ...807..132M} and the gravitational wave application therein. The results in Sections~\ref{Sec:PE_glitchers} and~\ref{Sec:PE_zeroglitchers} show that there is minimal posterior support for $\gamma=1$ , with greater than 99.99\% of the probability weight at $\gamma \leq 1$.  
The analysis in this paper argues strongly against $\lambda \propto \tau^{-1}$. 

The subset of non-glitching pulsars used here is incomplete, as noted in Section~\ref{Sec:nonglitchers}. Future work will include building up a larger set of non-glitching pulsars that are routinely monitored.  Moreover, the observation times $\Tobs$ used in this analysis may be overestimated.  Section~\ref{Sec:TobsBiases} examines this potential bias.
We emphasize that the $\Ng=0$ objects are included under the assumption that they are physically capable of glitching but have not done so to date. It is entirely conceivable that a subset of pulsars exist, that never glitch for some unknown physical reason, e.g. an internal stress threshold which is not surpassed. In this scenario, it would be misleading to lump together the $\Ng=0$ objects with the $\Ng\geq 1$ objects in the analyses in Sections~\ref{Sec:PE_glitchers} and~\ref{Sec:PE_zeroglitchers}.

Glitch rate laws can lend insight into neutron star interiors. For example, if pulsar glitches are caused by angular moment transfer between the superfluid core and rigid crust due to vortex unpinning, the rate-age relation may indicate how the abundance and depth of pinning potentials in the star evolve with age, whether they involve nuclear lattice sites~\citep{2020A&A...636A.101P,2016MNRAS.455.3952S} or magnetic fluxoids~\citep{1991PhR...203....1B,2017MNRAS.472.4851D,2018MNRAS.475..910D} or something else entirely.  
Model A is preferred across the whole population, without selecting by glitch size, possibly indicating that glitches are triggered when the \textit{fractional} crust-superfluid lag exceeds a threshold, as explained in Section~\ref{Sec:modelMotivation} (although other scenarios are certainly viable too). In contrast,~\citet{2017A&A...608A.131F} showed that when binning in $\nudot$, larger glitches with $\Delta\nu \geq 10\mu{\rm Hz}$ are primarily responsible for a linear relation between glitch activity and $\dot{\nu}$, with $\lambda = (4.2\pm 0.5) \times 10^2 \, {\rm Hz^{-1}} \, |\dot{\nu} |$ in Section 4.2 and Figure 6 of the latter reference. Physically this may indicate that larger glitches are triggered when the \textit{absolute} crust-superfluid lag exceeds a threshold. Disentangling these and other possibilities is an interesting focus for future work; the findings are not contradictory, because it is entirely plausible that small and large glitches, or objects with small and large $| \dot{\nu} |$, adhere to different rate laws.
The results may also point to a unified model for glitch activity in millisecond pulsars and younger pulsars.  Two glitches have been observed in millisecond pulsars over the last 50 years~\citep{McKee2016,2004ApJ...612L.125C}. While some work proposes that millisecond pulsars belong to a different population and are prevented from glitching by some underlying physical mechanism~\citep{Mandal2006,McKee2016}, they may also be compatible with a low but nonzero glitch rate, consistent with the general argument in Section 4.2 of ~\citet{2015ApJ...807..132M}, but with the specific form of the rate law proposed
($\lambda \propto \tau^{-1}$) disfavored strongly by the Bayesian analysis here in Sections~\ref{Sec:GlitchResults} and~\ref{Sec:NonglitchResults}.

Longer observations of pulsars and discoveries of new pulsars and glitches will refine this analysis, and so will better estimates of $\Tobs$. We strongly encourage future publications and surveys to report detailed and complete information about observing times and cadences. This information is sometimes hard to find in existing public data, yet it is critical for quantifying and interpreting statistical biases, including those discussed partially in Section~\ref{Sec:Biases}. 

Previous work hints that pulsar activity peaks at $\tau\sim10^{4}$ years~\citep{2000MNRAS.315..534L}, with  younger and older pulsars glitching more rarely.  Future work could devise a model to account for such a peak, as well as perform model selection to test statistically for its veracity.

We emphasize in closing that the analysis in this paper is not fundamentally new; it builds incrementally on extensive previous work~\citep{McKenna1990, 2000MNRAS.315..534L,Espinoza2011,2017A&A...608A.131F}. By its nature,
the question of the glitch rate law can only be answered gradually over time, with new analyses covering the same ground as old analyses and updating them, as new data become available, or new selection effects become noticed. The new incremental elements in this paper are: (i) formal model selection between Models A--D within a Bayesian framework; (ii) the inclusion of a large set of objects with $N_{\rm g}=0$, and (iii) an introductory study of $\Tobs$ selection effects, which builds on previous work. Even with these incremental steps forward, the glitch rate law is far from settled. The glitch microphysics remains unknown, so step-by-step testing of phenomenological models like Models A--D and many others will continue for the foreseeable future.

\section*{Acknowledgments}
We thank the anonymous referee, who pointed out several important references, suggested studying Model C (see Sections~\ref{Sec:Models}--\ref{Sec:NonglitchResults}) and observation time biases (see Section~\ref{Sec:Biases}), and improved the structure of the manuscript overall.
Parts of this research are supported by the Australian Research Council (ARC) Centre of Excellence for Gravitational
Wave Discovery (OzGrav) (project number CE170100004)
and ARC Discovery Project DP170103625.
JBC is supported by an Australian Postgraduate Award. LD is 
supported by an Australian Government Research Training Program 
Scholarship and by the Rowden White Scholarship.

\section*{Data Availability}
The data underlying this article are from the Jodrell Bank Glitch Catalog~\citep{Espinoza2011}, which can be accessed at \url{http://www.jb.man.ac.uk/pulsar/glitches.html}, and from the Australia Telescope National Facility~\citep{2005yCat.7245....0M,2005AJ....129.1993M} and can be accessed at \url{https://www.atnf.csiro.au/research/pulsar/psrcat/}.



\bibliographystyle{mnras}
\bibliography{glitch_populations}



\appendix
\section{Full posterior distributions}\label{Ap:Posteriors}
\begin{figure}
  \scalebox{0.5}{\includegraphics{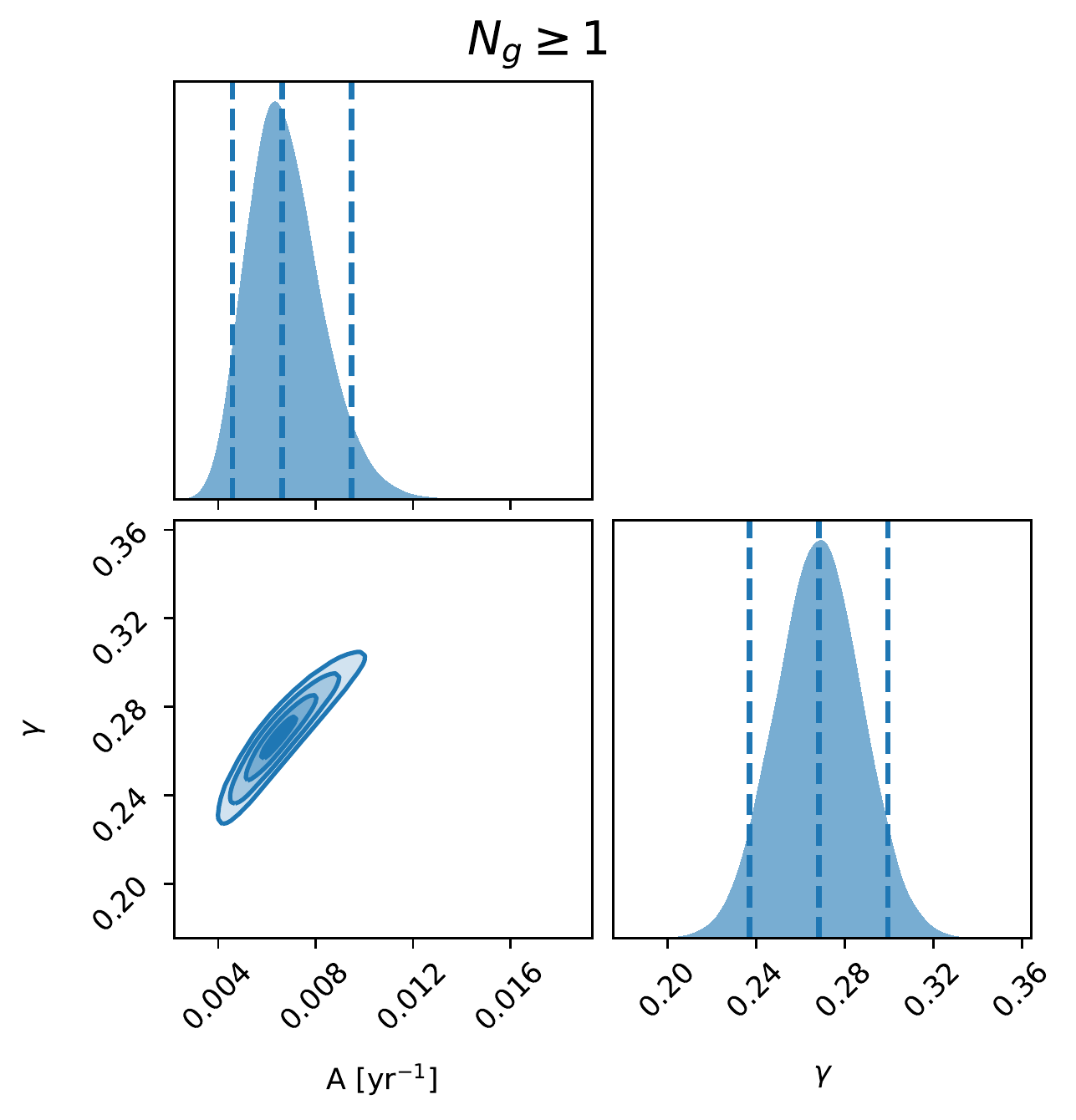}}
  \scalebox{0.5}{\includegraphics{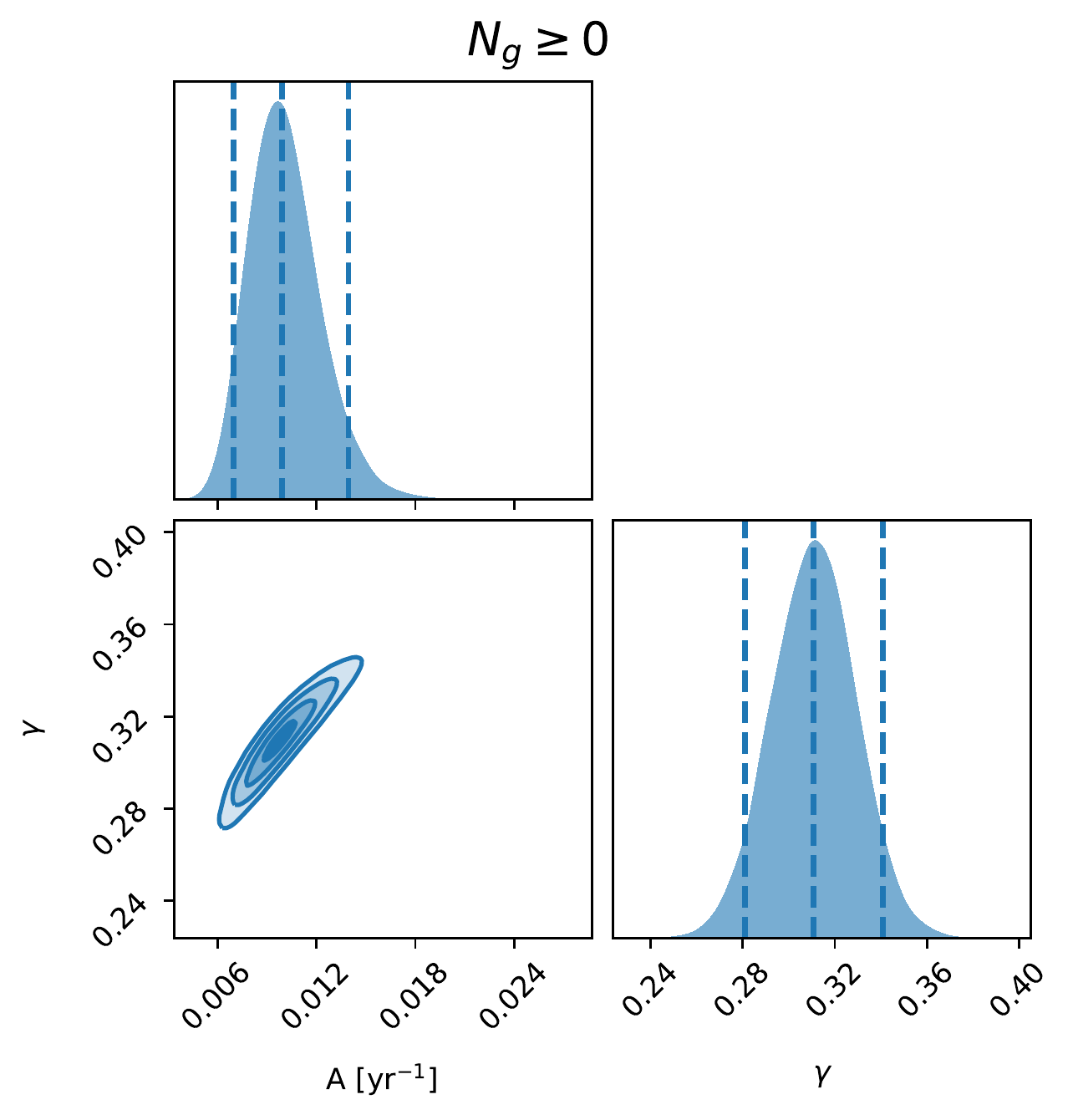}}
  \caption{Posterior distributions for Model A with parameters $A$ and $\gamma$. Left: $\Ng\geq1$. Right: $\Ng\geq0$.}
  \label{Fig:ModelAPE}
\end{figure}

\begin{figure}
  \scalebox{0.45}{\includegraphics{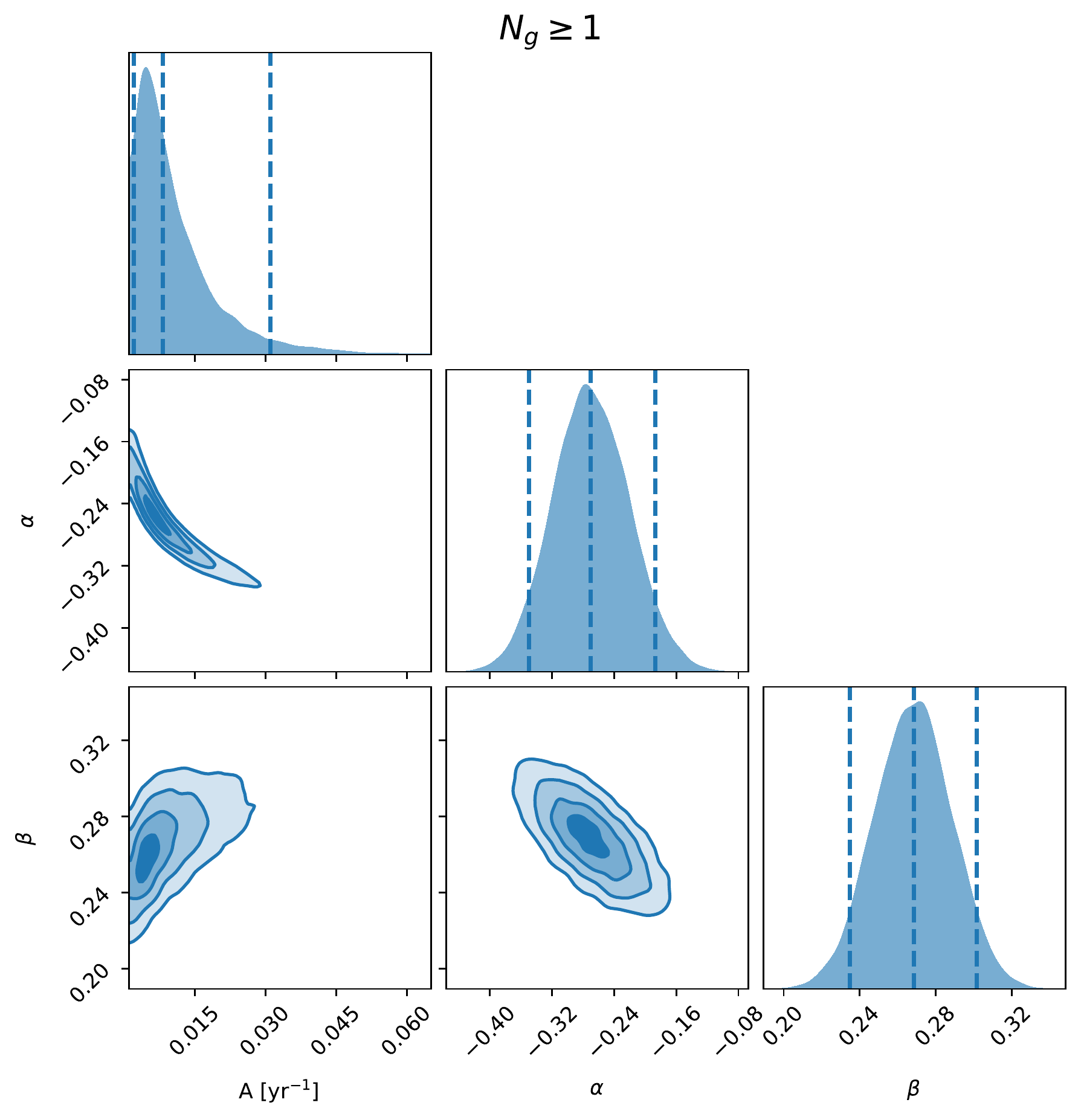}}
  \scalebox{0.45}{\includegraphics{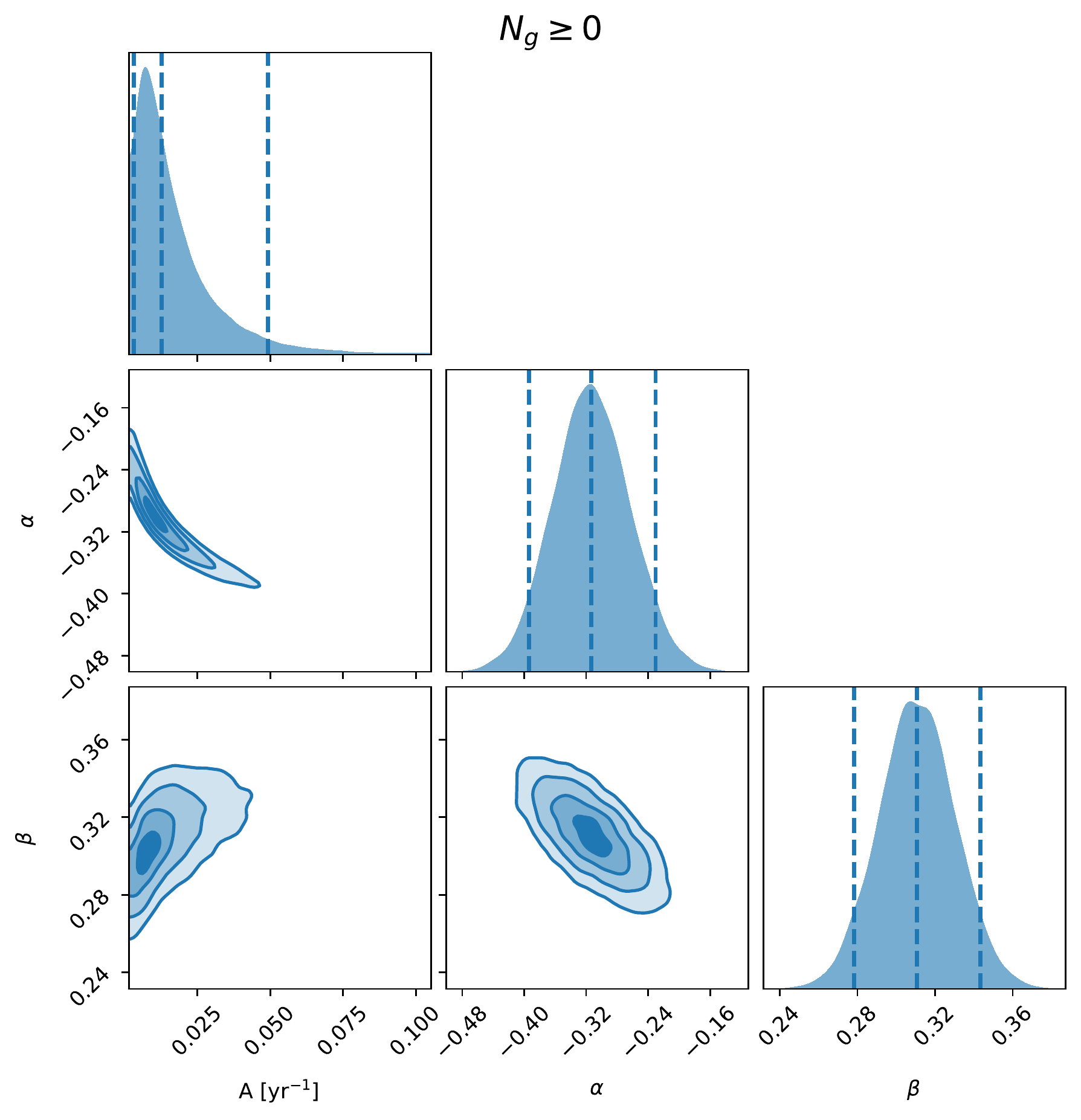}}
  \caption{Same as Figure~\ref{Fig:ModelAPE} but for Model B with parameters $A$, $\alpha$, and $\beta$.}
  \label{Fig:ModelBPE}
\end{figure}

\begin{figure}
  \scalebox{0.5}{\includegraphics{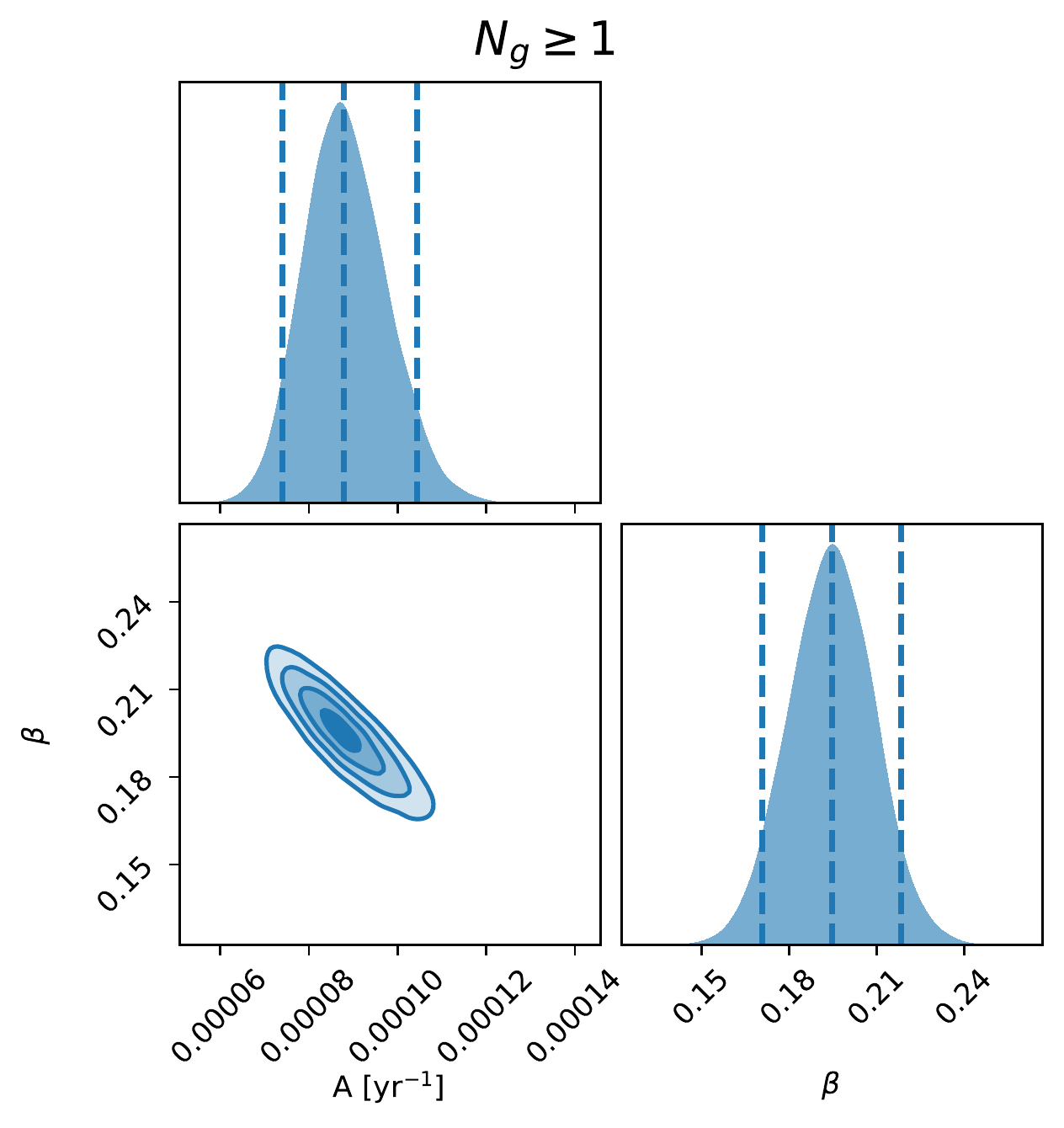}}
  \scalebox{0.5}{\includegraphics{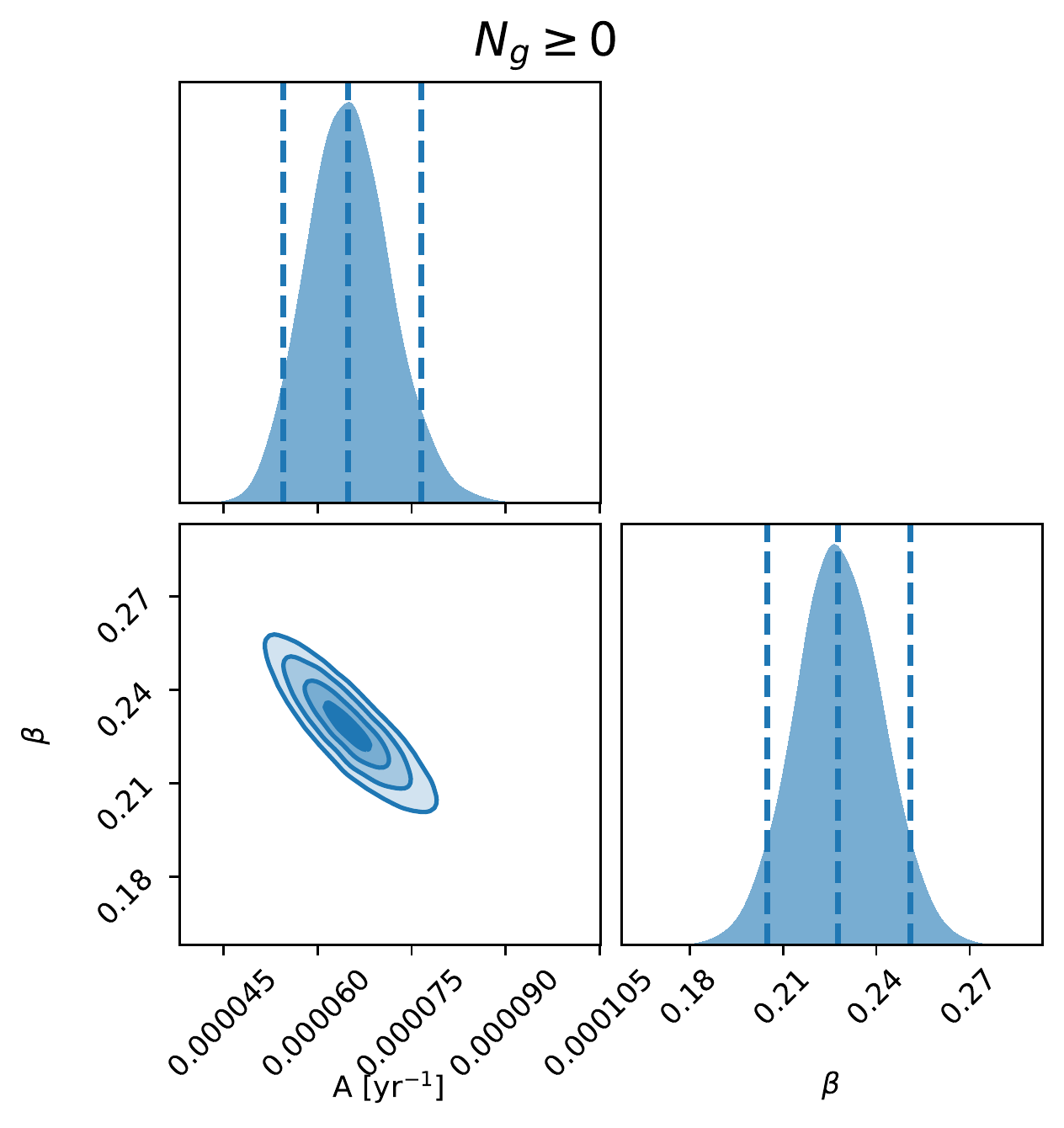}}
  \caption{Same as Figure~\ref{Fig:ModelAPE} but for Model C with parameter $A$ and $\beta$.}
  \label{Fig:ModelCPE}
\end{figure}

\begin{figure}
  \scalebox{0.5}{\includegraphics{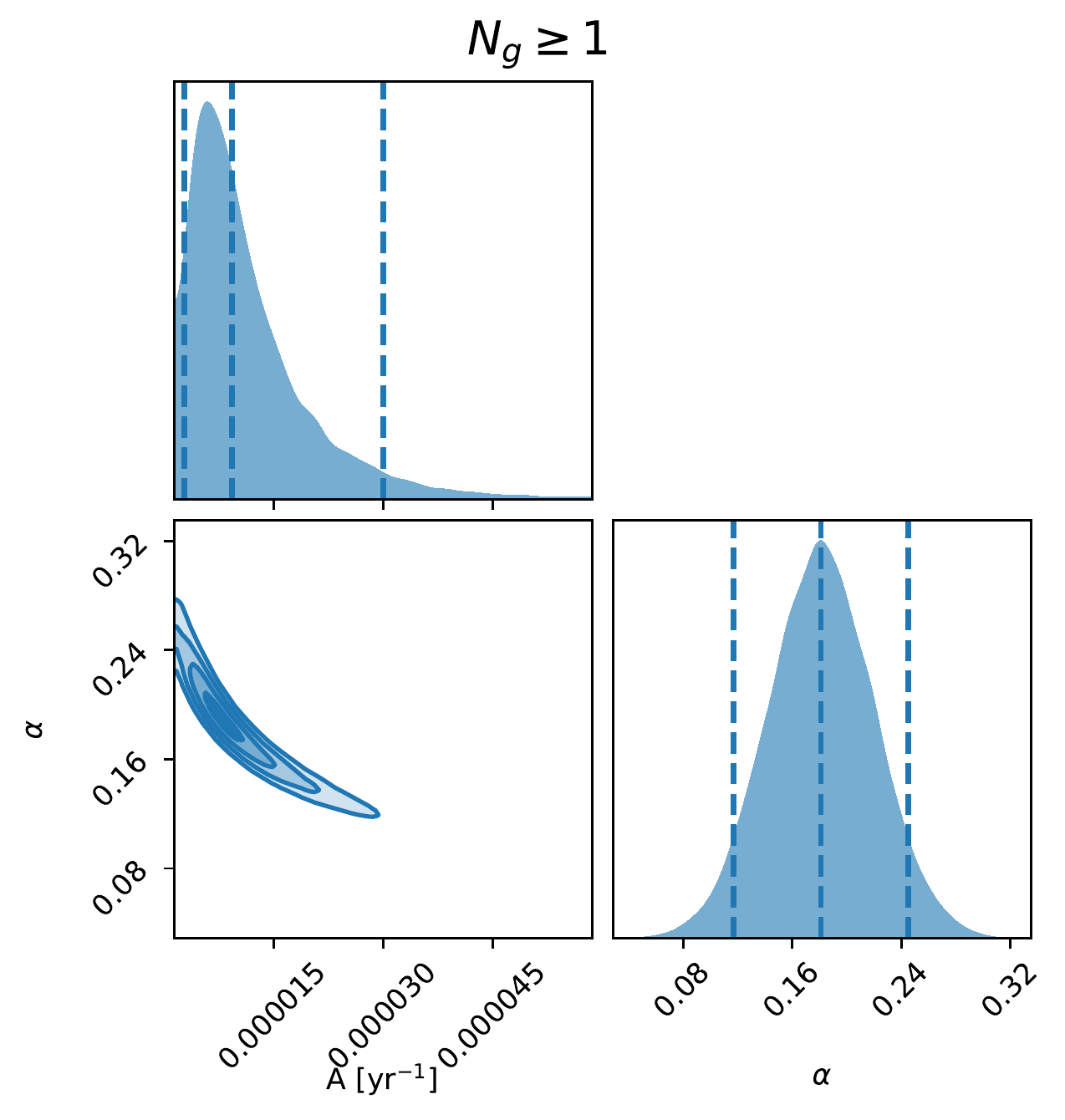}}
  \scalebox{0.5}{\includegraphics{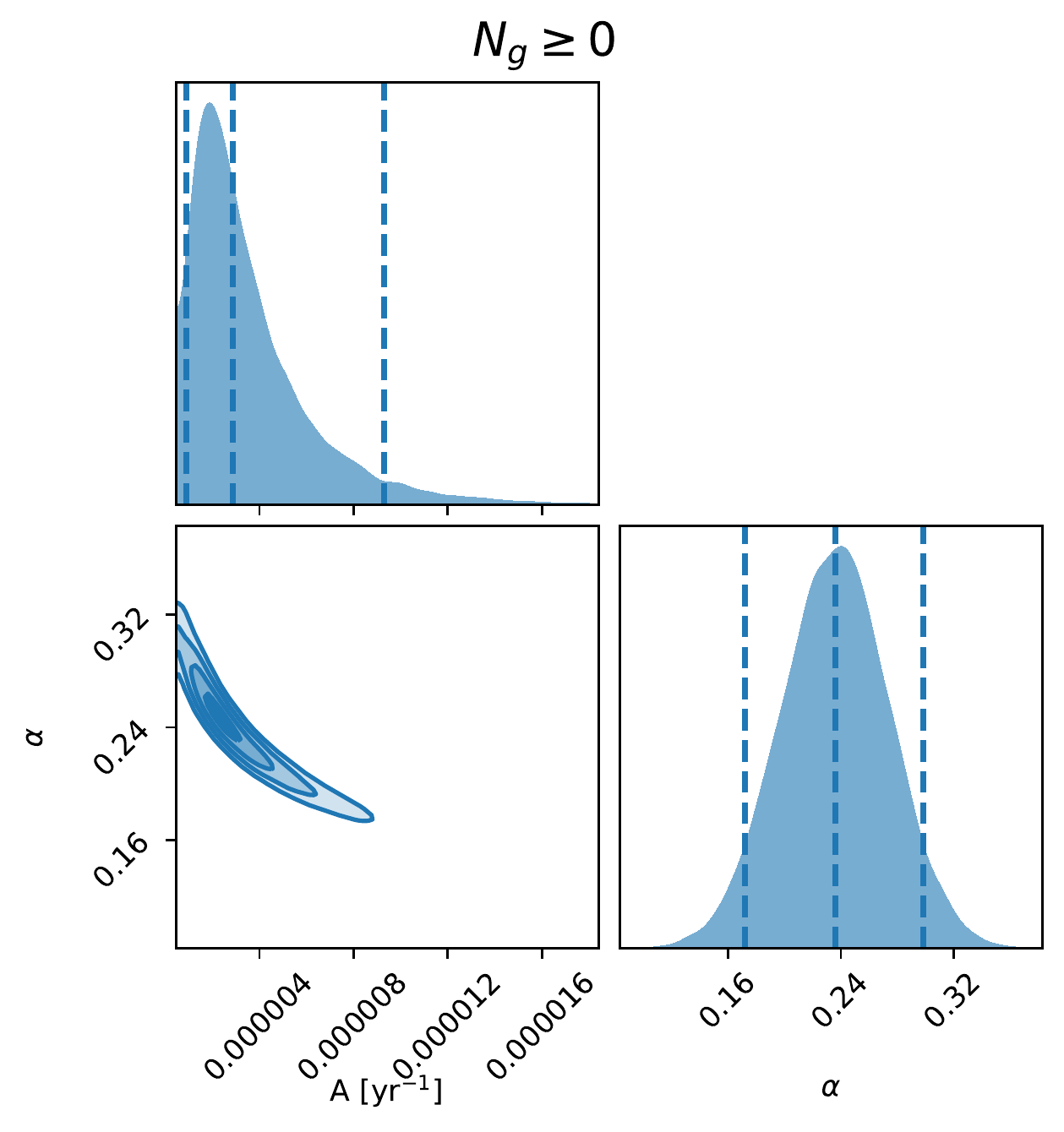}}
  \caption{Same as Figure~\ref{Fig:ModelAPE} but for Model D with parameters $A$ and $\alpha$.}
  \label{Fig:ModelDPE}
\end{figure}

Here we present plots of the full posterior distributions for Models A--D in Figures~\ref{Fig:ModelAPE}-\ref{Fig:ModelDPE}. In each figure, the left plot shows the results for the $\Ng\geq1$ sample, and the right plot shows the results for the $\Ng\geq0$ sample. We show the one-dimensional posteriors for each parameter, and the joint two-dimensional posterior. All posteriors are unimodal.

\section{List of pulsars with $\Ng>1$ used in this work}\label{Ap:GlitchTable}
Details of the objects with $\Ng>1$ are given in Table~\ref{Tab:rates}.

\bgroup
\def\arraystretch{1.5}
 \begin{table*}
\caption{List of pulsars with $\Ng\geq1$ used is this work.  For every pulsar we report the number of glitches, age, observation time,  and median estimate of the Poisson rate $\lambda$ with the upper and lower bounds of the $90\%$ credible interval. }
\begin{center}
\begin{tabular}{c c c c c c c c c c }
PSR & $N_\mathrm{glitch}$ & $\tau$ [kyr] & $T_\mathrm{obs}$ [days] & $\lambda$ [yr$^{-1}$] & PSR & $N_\mathrm{glitch}$ & $\tau$ [kyr] & $T_\mathrm{obs}$ [days] & $\lambda$ [yr$^{-1}$] \\ 
\hline
\hline
 J0007+7303 & 2 & 13.9 & 3956 & $0.154^{+0.12}_{-0.41}$ &  J0146+6145 & 2 & 69.1 & 9435 & $0.0655^{+0.052}_{-0.17}$ \\
 J0147+5922 & 1 & 12100 & 12722 & $0.0209^{+0.019}_{-0.084}$ &  J0157+6212 & 2 & 197 & 17471 & $0.0351^{+0.028}_{-0.09}$ \\
 J0205+6449 & 13 & 5.37 & 6513 & $0.711^{+0.28}_{-0.66}$ &  J0215+6218 & 1 & 13100 & 7974 & $0.0319^{+0.029}_{-0.13}$ \\
 J0358+5413 & 2 & 564 & 17471 & $0.035^{+0.028}_{-0.092}$ &  J0406+6138 & 1 & 1690 & 15279 & $0.0171^{+0.015}_{-0.07}$ \\
 J0415+6954 & 1 & 80800 & 12784 & $0.0878^{+0.081}_{-0.37}$ &
 J0502+4654 & 1 & 1810 & 15279 & $0.0169^{+0.015}_{-0.07}$ \\  J0525+1115 & 1 & 76300 & 15279 & $0.017^{+0.015}_{-0.07}$ &
 J0528+2200 & 4 & 1480 & 18932 & $0.0706^{+0.044}_{-0.12}$ \\  J0534+2200 & 30 & 1.26 & 18932 & $0.571^{+0.15}_{-0.34}$ &
 J0540-6919 & 1 & 1.67 & 13088 & $0.0203^{+0.018}_{-0.083}$ \\ J0601-0527 & 2 & 4820 & 15279 & $0.0406^{+0.032}_{-0.1}$ &
 J0611+1436 & 1 & 1070 & 1034 & $0.249^{+0.23}_{-1.1}$  \\ J0613-0200 & 1 & 5.06e+06 & 9070 & $0.0285^{+0.026}_{-0.12}$ & 
 J0631+1036 & 17 & 43.6 & 8705 & $0.702^{+0.25}_{-0.57}$ \\ J0633+1746 & 1 & 342 & 10166 & $0.0254^{+0.023}_{-0.11}$ &
 J0659+1414 & 2 & 111 & 15279 & $0.0401^{+0.032}_{-0.1}$ \\ J0729-1448 & 6 & 35.2 & 6513 & $0.317^{+0.17}_{-0.44}$ &
 J0729-1836 & 2 & 426 & 15279 & $0.0401^{+0.032}_{-0.11}$ \\ J0742-2822 & 8 & 157 & 17105 & $0.163^{+0.078}_{-0.19}$ & 
 J0758-1528 & 1 & 6680 & 15279 & $0.0171^{+0.015}_{-0.071}$ \\ J0834-4159 & 1 & 448 & 6148 & $0.041^{+0.037}_{-0.17}$ &
 J0846-3533 & 1 & 11000 & 15279 & $0.0173^{+0.016}_{-0.072}$ \\  J0855-3331 & 1 & 3180 & 15279 & $0.017^{+0.015}_{-0.07}$ &
 J0905-5127 & 2 & 221 & 8705 & $0.0713^{+0.057}_{-0.19}$ \\  J0922+0638 & 1 & 497 & 15279 & $0.017^{+0.015}_{-0.069}$ &
 J1016-5857 & 2 & 21 & 6878 & $0.0892^{+0.07}_{-0.23}$ \\  J1023-5746 & 1 & 4.6 & 3591 & $0.0699^{+0.064}_{-0.3}$ &
 J1048-5832 & 6 & 20.4 & 10166 & $0.203^{+0.11}_{-0.29}$ \\  J1052-5954 & 1 & 143 & 6148 & $0.0411^{+0.038}_{-0.17}$ &
 J1105-6107 & 5 & 63.2 & 8339 & $0.205^{+0.12}_{-0.32}$ \\  J1112-6103 & 2 & 32.7 & 6878 & $0.0894^{+0.07}_{-0.23}$ &
 J1119-6127 & 4 & 1.61 & 7244 & $0.186^{+0.12}_{-0.32}$ \\  J1123-6259 & 1 & 818 & 7974 & $0.0323^{+0.029}_{-0.13}$ &
 J1124-5916 & 1 & 2.85 & 6513 & $0.0391^{+0.036}_{-0.17}$ \\  J1141-3322 & 1 & 9960 & 7974 & $0.0323^{+0.03}_{-0.14}$ &
 J1301-6305 & 2 & 11 & 6878 & $0.0902^{+0.071}_{-0.23}$ \\  J1302-6350 & 1 & 332 & 10166 & $0.0255^{+0.023}_{-0.1}$ &
 J1328-4357 & 1 & 2800 & 15279 & $0.017^{+0.015}_{-0.07}$ \\ J1357-6429 & 2 & 7.31 & 5052 & $0.125^{+0.098}_{-0.32}$ & 
 J1410-6132 & 1 & 24.8 & 4322 & $0.0594^{+0.055}_{-0.25}$ \\ J1412-6145 & 1 & 50.4 & 6878 & $0.0368^{+0.034}_{-0.16}$ &
 J1413-6141 & 7 & 13.6 & 6148 & $0.394^{+0.2}_{-0.51}$  \\ J1413-6205 & 1 & 62.8 & 3591 & $0.0698^{+0.064}_{-0.3}$ &
 J1420-6048 & 5 & 13 & 6878 & $0.248^{+0.14}_{-0.38}$ \\
 J1452-6036 & 1 & 1690 & 6148 & $0.0408^{+0.037}_{-0.18}$ &  J1453-6413 & 1 & 1040 & 18566 & $0.0142^{+0.013}_{-0.057}$ \\
 J1509+5531 & 1 & 2340 & 18932 & $0.0137^{+0.012}_{-0.057}$ &  J1531-5610 & 1 & 96.7 & 6148 & $0.0411^{+0.038}_{-0.17}$ \\
 J1532+2745 & 1 & 22900 & 15279 & $0.0169^{+0.015}_{-0.07}$ &  J1539-5626 & 1 & 795 & 10166 & $0.0253^{+0.023}_{-0.11}$ \\
 J1614-5048 & 2 & 7.42 & 10166 & $0.0604^{+0.048}_{-0.16}$ &  J1617-5055 & 1 & 8.13 & 7974 & $0.0322^{+0.03}_{-0.13}$ \\
 J1644-4559 & 3 & 359 & 17105 & $0.0571^{+0.04}_{-0.12}$ &  J1646-4346 & 1 & 32.5 & 10166 & $0.0255^{+0.023}_{-0.11}$ \\
 J1647-4552 & 1 & 202 & 5052 & $0.0497^{+0.046}_{-0.21}$ &  J1702-4310 & 1 & 17 & 6148 & $0.0413^{+0.038}_{-0.17}$ \\
 J1705-1906 & 1 & 1140 & 15279 & $0.0169^{+0.015}_{-0.07}$ &  J1705-3423 & 3 & 3760 & 8705 & $0.112^{+0.078}_{-0.23}$ \\
 J1708-4008 & 6 & 8.9 & 8339 & $0.249^{+0.13}_{-0.35}$ &  J1709-1640 & 1 & 1640 & 18566 & $0.0141^{+0.013}_{-0.057}$ \\
 J1709-4429 & 5 & 17.5 & 10166 & $0.167^{+0.095}_{-0.26}$ &  J1718-3718 & 1 & 33.2 & 5783 & $0.0453^{+0.042}_{-0.18}$ \\
 J1718-3825 & 1 & 89.5 & 6878 & $0.0367^{+0.033}_{-0.15}$ &  J1720-1633 & 1 & 4280 & 12722 & $0.0202^{+0.018}_{-0.085}$ \\
 J1721-3532 & 1 & 176 & 10166 & $0.0251^{+0.023}_{-0.11}$ & J1726-3530 & 1 & 14.5 & 6878 & $0.0372^{+0.034}_{-0.16}$ \\
 J1730-3350 & 4 & 26 & 10166 & $0.133^{+0.084}_{-0.23}$ &  J1730-3353 & 1 & 2360 & 5783 & $0.0444^{+0.041}_{-0.19}$ \\
 J1731-4744 & 5 & 80.4 & 18932 & $0.09^{+0.052}_{-0.14}$ &  J1734-3333 & 1 & 8.13 & 6513 & $0.0395^{+0.036}_{-0.17}$ \\
 J1737-3102 & 1 & 326 & 6513 & $0.0386^{+0.035}_{-0.17}$ & J1737-3137 & 4 & 51.4 & 6513 & $0.205^{+0.13}_{-0.36}$ \\
 J1739-2903 & 1 & 650 & 12357 & $0.0209^{+0.019}_{-0.086}$ &  J1740+1000 & 2 & 114 & 7244 & $0.0829^{+0.065}_{-0.22}$ \\
 J1740-3015 & 36 & 20.6 & 12357 & $1.05^{+0.26}_{-0.57}$ &  J1743-3150 & 1 & 317 & 10166 & $0.0252^{+0.023}_{-0.1}$ \\
 J1746-2856 & 1 & 1200 & 5052 & $0.0499^{+0.046}_{-0.22}$ &  J1751-3323 & 4 & 984 & 6148 & $0.22^{+0.14}_{-0.38}$  \\
 J1755-0903 & 1 & 3870 & 2922 & $0.163^{+0.088}_{-0.23}$ &  J1757-2421 & 1 & 285 & 16740 & $0.0156^{+0.014}_{-0.064}$
  \end{tabular}
\end{center}
\label{Tab:rates}
\end{table*}

 \begin{table*}
\contcaption{}
\begin{center}
\begin{tabular}{c c c c c c c c c c }
PSR & $N_\mathrm{glitch}$ & $\tau$ [kyr] & $T_\mathrm{obs}$ [days] & $\lambda$ [yr $^{-1}$] & PSR & $N_\mathrm{glitch}$ & $\tau$ [kyr] & $T_\mathrm{obs}$ [days] & $\lambda$ [yr $^{-1}$] \\ 
\hline
\hline
 J1759-2922 & 1 & 1970 & 7974 & $0.0316^{+0.029}_{-0.14}$ &
 J1801-0357 & 1 & 4410 & 12722 & $0.02^{+0.018}_{-0.084}$ \\  J1801-2304 & 13 & 58.3 & 12722 & $0.363^{+0.14}_{-0.34}$ &
 J1801-2451 & 5 & 15.5 & 12722 & $0.135^{+0.077}_{-0.21}$ \\  J1803-2137 & 5 & 15.8 & 12357 & $0.138^{+0.08}_{-0.22}$ &
 J1806-2125 & 1 & 64.7 & 6513 & $0.0399^{+0.037}_{-0.17}$ \\  J1809-0119 & 2 & 5150 & 3591 & $0.173^{+0.13}_{-0.44}$ &
 J1809-1917 & 1 & 51.4 & 6513 & $0.0391^{+0.036}_{-0.16}$ \\  J1809-2004 & 2 & 946 & 6513 & $0.0942^{+0.074}_{-0.25}$ &
 J1812-1718 & 3 & 1000 & 12357 & $0.0797^{+0.056}_{-0.16}$ \\  J1813-1246 & 1 & 43.4 & 3956 & $0.0646^{+0.06}_{-0.27}$ &
 J1814-1744 & 7 & 84.6 & 7244 & $0.336^{+0.17}_{-0.43}$ \\  J1818-1422 & 1 & 2270 & 12357 & $0.0201^{+0.018}_{-0.085}$ &
 J1819-1458 & 2 & 120 & 5052 & $0.121^{+0.095}_{-0.32}$ \\  J1821-1419 & 2 & 29.3 & 5783 & $0.106^{+0.083}_{-0.27}$ &
 J1824-1118 & 1 & 1940 & 12357 & $0.0208^{+0.019}_{-0.088}$ \\  J1825-0935 & 7 & 233 & 17471 & $0.139^{+0.071}_{-0.18}$ &
 J1826-1334 & 6 & 21.4 & 12357 & $0.166^{+0.09}_{-0.23}$ \\  J1827-0958 & 1 & 3880 & 12357 & $0.0208^{+0.019}_{-0.087}$ &
 J1830-1059 & 1 & 107 & 11992 & $0.0214^{+0.019}_{-0.091}$ \\  J1830-1135 & 1 & 2060 & 6513 & $0.0398^{+0.036}_{-0.17}$ &
 J1832+0029 & 1 & 5600 & 5052 & $0.0507^{+0.047}_{-0.21}$ \\  J1833-0827 & 3 & 147 & 12357 & $0.0794^{+0.055}_{-0.16}$ &
 J1833-1034 & 4 & 4.85 & 5417 & $0.247^{+0.16}_{-0.43}$ \\  J1834-0731 & 1 & 140 & 5783 & $0.0435^{+0.04}_{-0.19}$ &
 J1835-1020 & 1 & 810 & 6513 & $0.0386^{+0.035}_{-0.16}$ \\  J1835-1106 & 2 & 128 & 8705 & $0.0706^{+0.056}_{-0.18}$ &
 J1836-1008 & 1 & 756 & 15279 & $0.0171^{+0.016}_{-0.071}$ \\  J1837-0559 & 2 & 963 & 6513 & $0.0941^{+0.074}_{-0.25}$ &
 J1837-0604 & 3 & 33.8 & 6878 & $0.143^{+0.099}_{-0.29}$ \\  J1838-0453 & 2 & 51.9 & 6513 & $0.0936^{+0.074}_{-0.25}$ &
 J1838-0537 & 1 & 4.89 & 2861 & $0.089^{+0.082}_{-0.38}$ \\  J1841-0157 & 1 & 581 & 5783 & $0.0438^{+0.04}_{-0.19}$ &
 J1841-0425 & 1 & 461 & 12357 & $0.0204^{+0.019}_{-0.087}$ \\  J1841-0456 & 3 & 4.57 & 8339 & $0.117^{+0.081}_{-0.24}$ &
 J1841-0524 & 5 & 30.2 & 5783 & $0.292^{+0.17}_{-0.46}$ \\  J1842+0257 & 1 & 1650 & 5052 & $0.0494^{+0.046}_{-0.21}$ &
 J1844-0310 & 1 & 813 & 7244 & $0.0355^{+0.033}_{-0.15}$ \\  J1844-0433 & 1 & 4010 & 11992 & $0.0211^{+0.019}_{-0.09}$ &
 J1844-0538 & 1 & 417 & 12357 & $0.0208^{+0.019}_{-0.088}$ \\  J1845-0316 & 2 & 371 & 7244 & $0.0847^{+0.067}_{-0.22}$ &
 J1846-0258 & 2 & 0.728 & 7244 & $0.0843^{+0.066}_{-0.22}$ \\  J1847-0130 & 2 & 83.3 & 6148 & $0.101^{+0.08}_{-0.26}$ &
 J1847-0402 & 2 & 183 & 18201 & $0.0338^{+0.027}_{-0.087}$ \\  J1850-0026 & 4 & 67.5 & 3956 & $0.338^{+0.21}_{-0.6}$ &
 J1851-0029 & 1 & 1730 & 3956 & $0.0649^{+0.06}_{-0.27}$ \\  J1853+0056 & 2 & 204 & 6513 & $0.0956^{+0.075}_{-0.24}$ &
 J1853+0545 & 1 & 3270 & 6148 & $0.0414^{+0.038}_{-0.17}$ \\  J1856+0113 & 2 & 20.3 & 10531 & $0.0585^{+0.046}_{-0.16}$ &
 J1856+0245 & 1 & 20.6 & 4322 & $0.0599^{+0.055}_{-0.25}$ \\  J1901+0156 & 1 & 1940 & 12722 & $0.0201^{+0.018}_{-0.084}$ &
 J1901+0716 & 1 & 4460 & 11992 & $0.0213^{+0.019}_{-0.088}$ \\  J1902+0615 & 6 & 1380 & 14183 & $0.146^{+0.079}_{-0.2}$ &
 J1907+0602 & 2 & 19.5 & 3956 & $0.156^{+0.12}_{-0.41}$ \\  J1909+0007 & 3 & 2920 & 17105 & $0.057^{+0.039}_{-0.12}$ &
 J1909+0749 & 1 & 24.7 & 2495 & $0.102^{+0.094}_{-0.43}$ \\  J1909+0912 & 1 & 98.7 & 6513 & $0.0388^{+0.035}_{-0.16}$ &
 J1909+1102 & 1 & 1700 & 17105 & $0.015^{+0.014}_{-0.063}$ \\  J1910+0358 & 1 & 8260 & 16375 & $0.0158^{+0.014}_{-0.067}$ &
 J1910-0309 & 3 & 3650 & 15279 & $0.0643^{+0.045}_{-0.13}$ \\  J1913+0446 & 3 & 91.8 & 6148 & $0.158^{+0.11}_{-0.32}$ &
 J1913+0832 & 1 & 466 & 6513 & $0.0394^{+0.036}_{-0.17}$ \\  J1913+0904 & 1 & 147 & 5052 & $0.0505^{+0.046}_{-0.21}$ &
 J1913+1000 & 1 & 792 & 5783 & $0.0434^{+0.04}_{-0.19}$ \\  J1913+1011 & 1 & 169 & 6513 & $0.0392^{+0.036}_{-0.16}$ &
 J1915+1009 & 1 & 420 & 16375 & $0.0158^{+0.014}_{-0.066}$ \\  J1915+1606 & 1 & 109000 & 16375 & $0.0159^{+0.014}_{-0.066}$ &
 J1919+0021 & 1 & 2630 & 17471 & $0.0148^{+0.013}_{-0.063}$ \\  J1921+0812 & 1 & 622 & 5052 & $0.0507^{+0.047}_{-0.22}$ &
 J1926+0431 & 1 & 6920 & 15279 & $0.0169^{+0.015}_{-0.071}$ \\  J1932+2220 & 3 & 39.8 & 16375 & $0.0595^{+0.041}_{-0.12}$ &
 J1937+2544 & 1 & 4950 & 12722 & $0.0206^{+0.019}_{-0.083}$ \\  J1949-2524 & 1 & 4640 & 15279 & $0.0168^{+0.015}_{-0.071}$ &
 J1952+3252 & 6 & 107 & 11627 & $0.18^{+0.096}_{-0.25}$ \\  J1955+5059 & 2 & 5990 & 15279 & $0.0403^{+0.032}_{-0.11}$ &
  J1957+2831 & 4 & 1570 & 7974 & $0.0296^{+0.027}_{-0.12}$ \\  J2005-0020 & 1 & 1410 & 8705 & $0.207^{+0.13}_{-0.36}$ &
 J2021+3651 & 4 & 17.2 & 6513 & $0.166^{+0.1}_{-0.3}$ \\  J2022+2854 & 1 & 2870 & 17105 & $0.078^{+0.072}_{-0.33}$ &
 J2022+3842 & 1 & 8.94 & 3226 & $0.0203^{+0.019}_{-0.084}$ \\  J2029+3744 & 1 & 1560 & 12722 & $0.0151^{+0.014}_{-0.063}$ &
 J2032+4127 & 1 & 201 & 3956 & $0.0169^{+0.015}_{-0.071}$
 \end{tabular}
\end{center}
\end{table*}%

\begin{table*}
\contcaption{}
\begin{center}
\begin{tabular}{c c c c c c c c c c }
PSR & $N_\mathrm{glitch}$ & $\tau$ [kyr] & $T_\mathrm{obs}$ [days] & $\lambda$ [yr $^{-1/}$] & PSR & $N_\mathrm{glitch}$ & $\tau$ [kyr] & $T_\mathrm{obs}$ [days] & $\lambda$ [yr $^{-1/}$] \\ 
\hline
\hline
 J2116+1414 & 1 & 24100 & 15279 & $0.014^{+0.013}_{-0.059}$ &
 J2219+4754 & 1 & 3090 & 18566 & $0.0648^{+0.06}_{-0.27}$ \\  J2225+6535 & 5 & 1120 & 17105 & $0.302^{+0.16}_{-0.41}$ &
 J2229+6114 & 6 & 10.5 & 6878 & $0.0146^{+0.013}_{-0.061}$ \\  J2257+5909 & 1 & 1010 & 17471 & $0.101^{+0.058}_{-0.15}$ &
 J2301+5852 & 4 & 235 & 14183 & $0.0202^{+0.018}_{-0.083}$ \\  J2337+6151 & 1 & 40.6 & 12722 & $0.0296^{+0.027}_{-0.12}$ &
 J2346-0609 & 1 & 13700 & 8705 & $0.0945^{+0.059}_{-0.16}$ \\
 \hline
 \end{tabular}
\end{center}
\end{table*}%
\egroup

\section{List of pulsars with $\Ng=0$ used in this work}\label{Ap:ZeroGlitchTable}
Details of the objects with $\Ng=0$ are given in Table~\ref{Tab:zero_glitchers}.
\bgroup
\def\arraystretch{1.5}
 \begin{table*}
\caption{List of pulsars with $\Ng=0$ used is this work.  For every pulsar we report the characteristic spin-down age $\tau$ and observation time $\Tobs$. }
\begin{center}
\begin{tabular}{c c c c c c c c c  }
PSR & $\tau$ [kyr] & $T_\mathrm{obs}$ [days] & PSR & $\tau$ [kyr] & $T_\mathrm{obs}$ [days] & PSR & $\tau$ [kyr] & $T_\mathrm{obs}$ [days] \\
\hline
\hline
J0030+0451 & $7.58\times 10^6$ & 1553 & J0134-2937 & $2.77\times 10^4$ & 1553 & J0151-0635 & $5.24\times 10^4$ & 1553  \\ 
J0152-1637 & $1.02\times 10^4$ & 1553 & J0206-4028 & $8.35\times 10^3$ & 1553 & J0255-5304 & $2.27\times 10^5$ & 1553  \\ 
J0401-7608 & $5.6\times 10^3$ & 1553 & J0418-4154 & $9.09\times 10^3$ & 1553 & J0450-1248 & $6.76\times 10^4$ & 1553  \\ 
J0452-1759 & $1.51\times 10^3$ & 1553 & J0533+0402 & $9.54\times 10^4$ & 1553 & J0536-7543 & $3.42\times 10^4$ & 1553  \\ 
J0624-0424 & $1.98\times 10^4$ & 1553 & J0627+0706 & $253$ & 1553 & J0630-2834 & $2.77\times 10^3$ & 1553  \\ 
J0646+0905 & $1.95\times 10^4$ & 1553 & J0711-6830 & $5.84\times 10^6$ & 1553 & J0738-4042 & $4.32\times 10^3$ & 1553  \\ 
J0809-4753 & $2.82\times 10^3$ & 1553 & J0820-1350 & $9.32\times 10^3$ & 1553 & J0820-4114 & $3.87\times 10^5$ & 1553  \\ 
J0837+0610 & $2.97\times 10^3$ & 1553 & J0837-4135 & $3.37\times 10^3$ & 1553 & J0840-5332 & $6.98\times 10^3$ & 1553  \\ 
J0842-4851 & $1.07\times 10^3$ & 1553 & J0856-6137 & $9.09\times 10^3$ & 1553 & J0904-4246 & $8.1\times 10^3$ & 1553  \\ 
J0904-7459 & $1.88\times 10^4$ & 1553 & J0907-5157 & $2.19\times 10^3$ & 1553 & J0908-1739 & $9.5\times 10^3$ & 1553  \\ 
J0909-7212 & $6.48\times 10^4$ & 1553 & J0924-5302 & $333$ & 1553 & J0924-5814 & $2.38\times 10^3$ & 1553  \\ 
J0934-5249 & $4.92\times 10^3$ & 1553 & J0942-5552 & $461$ & 1553 & J0942-5657 & $323$ & 1553  \\ 
J0944-1354 & $2\times 10^5$ & 1553 & J0953+0755 & $1.75\times 10^4$ & 1553 & J0955-5304 & $3.87\times 10^3$ & 1553  \\ 
J0959-4809 & $1.29\times 10^5$ & 1553 & J1001-5507 & $441$ & 1553 & J1003-4747 & $2.35\times 10^3$ & 1553  \\ 
J1012-5857 & $730$ & 1553 & J1013-5934 & $1.26\times 10^4$ & 1553 & J1016-5345 & $6.33\times 10^3$ & 1553  \\ 
J1017-5621 & $2.54\times 10^3$ & 1553 & J1032-5911 & $2.45\times 10^3$ & 1553 & J1034-3224 & $7.91\times 10^4$ & 1553  \\ 
J1036-4926 & $4.9\times 10^3$ & 1553 & J1041-1942 & $2.32\times 10^4$ & 1553 & J1042-5521 & $2.76\times 10^3$ & 1553  \\ 
J1043-6116 & $439$ & 1553 & J1046-5813 & $5.11\times 10^3$ & 1553 & J1047-6709 & $1.86\times 10^3$ & 1553  \\ 
J1056-6258 & $1.87\times 10^3$ & 1553 & J1057-7914 & $1.61\times 10^4$ & 1553 & J1059-5742 & $4.37\times 10^3$ & 1553  \\ 
J1110-5637 & $4.3\times 10^3$ & 1553 & J1112-6613 & $6.43\times 10^3$ & 1553 & J1112-6926 & $4.58\times 10^3$ & 1553  \\ 
J1114-6100 & $303$ & 1553 & J1116-4122 & $1.88\times 10^3$ & 1553 & J1121-5444 & $3.05\times 10^3$ & 1553  \\ 
J1126-6942 & $2.78\times 10^3$ & 1553 & J1133-6250 & $3.59\times 10^4$ & 1553 & J1136+1551 & $5.04\times 10^3$ & 1553  \\ 
J1136-5525 & $702$ & 1553 & J1146-6030 & $2.42\times 10^3$ & 1553 & J1157-6224 & $1.61\times 10^3$ & 1553  \\ 
J1202-5820 & $3.37\times 10^3$ & 1553 & J1210-5559 & $6.11\times 10^3$ & 1553 & J1224-6407 & $692$ & 1553  \\ 
J1231-6303 & $1.53\times 10^4$ & 1553 & J1239-6832 & $1.73\times 10^3$ & 1553 & J1243-6423 & $1.37\times 10^3$ & 1553  \\ 
J1253-5820 & $1.93\times 10^3$ & 1553 & J1259-6741 & $8.71\times 10^3$ & 1553 & J1305-6455 & $2.24\times 10^3$ & 1553  \\ 
J1306-6617 & $1.25\times 10^3$ & 1553 & J1312-5402 & $8.06\times 10^4$ & 1553 & J1312-5516 & $2.36\times 10^3$ & 1553  \\ 
J1320-5359 & $479$ & 1553 & J1326-5859 & $2.34\times 10^3$ & 1553 & J1326-6408 & $4.06\times 10^3$ & 1553  \\ 
J1326-6700 & $1.62\times 10^3$ & 1553 & J1327-6222 & $447$ & 1553 & J1327-6301 & $2.04\times 10^3$ & 1553  \\ 
J1338-6204 & $1.42\times 10^3$ & 1553 & J1350-5115 & $6.23\times 10^3$ & 1553 & J1355-5153 & $3.63\times 10^3$ & 1553  \\ 
J1356-5521 & $1.11\times 10^4$ & 1553 & J1401-6357 & $791$ & 1553 & J1418-3921 & $1.95\times 10^4$ & 1553  \\ 
J1420-5416 & $6.39\times 10^4$ & 1553 & J1424-5822 & $1.47\times 10^3$ & 1553 & J1428-5530 & $4.33\times 10^3$ & 1553  \\ 
J1430-6623 & $4.48\times 10^3$ & 1553 & J1435-5954 & $4.86\times 10^3$ & 1553 & J1456-6843 & $4.22\times 10^4$ & 1553  \\ 
J1457-5122 & $5.24\times 10^3$ & 1553 & J1507-4352 & $2.87\times 10^3$ & 1553 & J1507-6640 & $4.89\times 10^3$ & 1553  \\ 
J1511-5414 & $6.55\times 10^3$ & 1553 & J1512-5759 & $298$ & 1553 & J1514-4834 & $7.79\times 10^3$ & 1553  \\ 
J1522-5829 & $3.05\times 10^3$ & 1553 & J1527-3931 & $2.01\times 10^3$ & 1553 & J1527-5552 & $1.47\times 10^3$ & 1553  \\ 
J1534-5334 & $1.52\times 10^4$ & 1553 & J1534-5405 & $2.97\times 10^3$ & 1553 & J1542-5034 & $2.39\times 10^3$ & 1553  \\ 
J1543+0929 & $2.74\times 10^4$ & 1553 & J1544-5308 & $4.69\times 10^4$ & 1553 & J1549-4848 & $324$ & 1553  \\ 
J1553-5456 & $1.09\times 10^3$ & 1553 & J1555-3134 & $1.32\times 10^5$ & 1553 & J1557-4258 & $1.58\times 10^4$ & 1553  \\ 
J1559-4438 & $4\times 10^3$ & 1553 & J1559-5545 & $741$ & 1553 & J1600-5044 & $603$ & 1553  \\ 
J1603-2531 & $2.82\times 10^3$ & 1553 & J1603-2712 & $4.1\times 10^3$ & 1553 & J1604-4909 & $5.09\times 10^3$ & 1553  \\ 
J1605-5257 & $4.07\times 10^4$ & 1553 & J1613-4714 & $9.55\times 10^3$ & 1553 & J1623-0908 & $7.84\times 10^3$ & 1553  \\ 
J1623-4256 & $5.65\times 10^3$ & 1553 & J1626-4537 & $708$ & 1553 & J1633-4453 & $1.12\times 10^3$ & 1553  \\ 
 \end{tabular}
\end{center}
\label{Tab:zero_glitchers}
\end{table*}%

 \begin{table*}
\contcaption{}
\begin{center}
\begin{tabular}{c c c c c c c c c  }
PSR & $\tau$ [kyr] & $T_\mathrm{obs}$  [days] & PSR & $\tau$ [kyr] & $T_\mathrm{obs}$ [days] & PSR & $\tau$ [kyr] & $T_\mathrm{obs}$ [days] \\
\hline
\hline
J1633-5015 & $1.47\times 10^3$ & 1553 & J1639-4604 & $1.45\times 10^3$ & 1553 & J1646-6831 & $1.66\times 10^4$ & 1553  \\ 
J1651-4246 & $2.81\times 10^3$ & 1553 & J1651-5222 & $5.55\times 10^3$ & 1553 & J1651-5255 & $6.82\times 10^3$ & 1553  \\ 
J1652-2404 & $8.55\times 10^3$ & 1553 & J1700-3312 & $4.57\times 10^3$ & 1553 & J1701-3726 & $3.5\times 10^3$ & 1553  \\ 
J1703-1846 & $7.36\times 10^3$ & 1553 & J1703-3241 & $2.91\times 10^4$ & 1553 & J1707-4053 & $4.78\times 10^3$ & 1553  \\ 
J1708-3426 & $2.61\times 10^3$ & 1553 & J1711-5350 & $920$ & 1553 & J1715-4034 & $1.08\times 10^4$ & 1553  \\ 
J1717-3425 & $1.06\times 10^3$ & 1553 & J1717-4054 & $3.8\times 10^3$ & 1553 & J1720-2933 & $1.32\times 10^4$ & 1553  \\ 
J1722-3207 & $1.17\times 10^4$ & 1553 & J1722-3712 & $344$ & 1553 & J1727-2739 & $1.86\times 10^4$ & 1553  \\ 
J1730-2304 & $6.38\times 10^6$ & 1553 & J1733-2228 & $3.23\times 10^5$ & 1553 & J1736-2457 & $1.22\times 10^4$ & 1553  \\ 
J1741-3927 & $4.74\times 10^3$ & 1553 & J1745-3040 & $546$ & 1553 & J1751-4657 & $9.06\times 10^3$ & 1553  \\ 
J1752-2806 & $1.1\times 10^3$ & 1553 & J1759-2205 & $672$ & 1553 & J1759-3107 & $4.54\times 10^3$ & 1553  \\ 
J1801-2920 & $5.21\times 10^3$ & 1553 & J1805-1504 & $6.84\times 10^4$ & 1553 & J1807-0847 & $9.01\times 10^4$ & 1553  \\ 
J1807-2715 & $1.08\times 10^3$ & 1553 & J1808-0813 & $1.12\times 10^4$ & 1553 & J1809-2109 & $2.91\times 10^3$ & 1553  \\ 
J1810-5338 & $1.08\times 10^4$ & 1553 & J1816-2650 & $1.41\times 10^5$ & 1553 & J1820-0427 & $1.5\times 10^3$ & 1553  \\ 
J1822-2256 & $2.19\times 10^4$ & 1553 & J1823-0154 & $1.06\times 10^4$ & 1553 & J1823-3106 & $1.54\times 10^3$ & 1553  \\ 
J1824-0127 & $1.01\times 10^4$ & 1553 & J1824-1945 & $573$ & 1553 & J1827-0750 & $2.77\times 10^3$ & 1553  \\ 
J1829-1751 & $877$ & 1553 & J1832-0827 & $161$ & 1553 & J1833-0338 & $262$ & 1553  \\ 
J1834-0426 & $6.39\times 10^4$ & 1553 & J1836-0436 & $3.38\times 10^3$ & 1553 & J1837-0653 & $3.91\times 10^4$ & 1553  \\ 
J1840-0809 & $6.44\times 10^3$ & 1553 & J1840-0815 & $7.19\times 10^3$ & 1553 & J1841+0912 & $5.54\times 10^3$ & 1553  \\ 
J1842-0359 & $5.73\times 10^4$ & 1553 & J1843-0000 & $1.79\times 10^3$ & 1553 & J1845-0743 & $4.52\times 10^3$ & 1553  \\ 
J1848-0123 & $1.99\times 10^3$ & 1553 & J1849-0636 & $497$ & 1553 & J1852-0635 & $568$ & 1553  \\ 
J1852-2610 & $6.08\times 10^4$ & 1553 & J1857+0212 & $164$ & 1553 & J1900-2600 & $4.74\times 10^4$ & 1553  \\ 
J1901+0331 & $1.39\times 10^3$ & 1553 & J1901-0906 & $1.72\times 10^4$ & 1553 & J1902+0556 & $919$ & 1553  \\ 
J1903+0135 & $2.87\times 10^3$ & 1553 & J1903-0632 & $2.01\times 10^3$ & 1553 & J1905-0056 & $3.33\times 10^3$ & 1553  \\ 
J1909+0254 & $2.84\times 10^3$ & 1553 & J1913-0440 & $3.22\times 10^3$ & 1553 & J1913+1400 & $1.03\times 10^4$ & 1553  \\ 
J1916+0951 & $1.7\times 10^3$ & 1553 & J1916+1312 & $1.22\times 10^3$ & 1553 & J1917+1353 & $428$ & 1553  \\ 
J1932+1059 & $3.1\times 10^3$ & 1553 & J1932-3655 & $3.19\times 10^4$ & 1553 & J1935+1616 & $947$ & 1553  \\ 
J1941-2602 & $6.68\times 10^3$ & 1553 & J1943-1237 & $9.31\times 10^3$ & 1553 & J1945-0040 & $3.1\times 10^4$ & 1553  \\ 
J1946-2913 & $1.02\times 10^4$ & 1553 & J2006-0807 & $2\times 10^5$ & 1553 & J2033+0042 & $8.19\times 10^3$ & 1553  \\ 
J2038-3816 & $6.05\times 10^3$ & 1553 & J2046-0421 & $1.67\times 10^4$ & 1553 & J2046+1540 & $9.89\times 10^4$ & 1553  \\ 
J2048-1616 & $2.84\times 10^3$ & 1553 & J2053-7200 & $2.74\times 10^4$ & 1553 & J2144-3933 & $2.72\times 10^5$ & 1553  \\ 
J2155-3118 & $1.32\times 10^4$ & 1553 & J2248-0101 & $1.15\times 10^4$ & 1553 & J2324-6054 & $1.44\times 10^4$ & 1553  \\
J2330-2005 & $5.62\times 10^3$ & 1553 &  - & - & - &  - & - & - \\
\hline
\end{tabular}
\end{center}
\end{table*}%
\egroup


\bsp	
\label{lastpage}
\end{document}